\documentclass[reprint,amsmath,amssymb, aps,pre,floatfix,superscriptaddress]{revtex4-1}

\usepackage{graphicx} 
\usepackage{siunitx}
\usepackage[version=4]{mhchem}
\usepackage{comment}
\usepackage{xcolor}
\usepackage{tikz}
\usepackage{scalerel}
\usepackage{xifthen}
\usepackage{chngcntr}
\usepackage{xcolor}
\usepackage{bm}
\usetikzlibrary{svg.path}
\usepackage[colorlinks,allcolors=blue]{hyperref}

\newcommand{\figref}[2][{}]{Fig.\ \ref{#2}\ifthenelse{\isempty{#1}}{}{\,(#1)}}
\newcommand{\eqnref}[1]{Eqn.~\ref{#1}}
\newcommand{\plb}[1]{(\MakeLowercase{#1})} 
\renewcommand{\vec}{\mathbf}

\DeclareMathOperator{\sgn}{sgn}
 \DeclareMathOperator\erf{erf}

\definecolor{orcidlogocol}{HTML}{A6CE39}
\tikzset{
  orcidlogo/.pic={
    \fill[orcidlogocol] svg{M256,128c0,70.7-57.3,128-128,128C57.3,256,0,198.7,0,128C0,57.3,57.3,0,128,0C198.7,0,256,57.3,256,128z};
    \fill[white] svg{M86.3,186.2H70.9V79.1h15.4v48.4V186.2z}
                 svg{M108.9,79.1h41.6c39.6,0,57,28.3,57,53.6c0,27.5-21.5,53.6-56.8,53.6h-41.8V79.1z M124.3,172.4h24.5c34.9,0,42.9-26.5,42.9-39.7c0-21.5-13.7-39.7-43.7-39.7h-23.7V172.4z}
                 svg{M88.7,56.8c0,5.5-4.5,10.1-10.1,10.1c-5.6,0-10.1-4.6-10.1-10.1c0-5.6,4.5-10.1,10.1-10.1C84.2,46.7,88.7,51.3,88.7,56.8z};
  }
}

\newcommand\orcid[1]{\href{https://orcid.org/#1}{\mbox{\scalerel*{
\begin{tikzpicture}[yscale=-1,transform shape]
\pic{orcidlogo};
\end{tikzpicture}
}{|}}}}

\begin{document}

\setlength{\unitlength}{1cm}

\newcommand{\goeaffil}{Max Planck Institute for Dynamics and Self-Organization, Am Fa\ss{}berg 17, 37077 G\"ottingen and Institute for the Dynamics of Complex Systems, Georg August Universit\"at G\"ottingen, Germany}
\newcommand{\oxaffil}{Rudolf Peierls Centre for Theoretical Physics, University of Oxford, Oxford OX1 3PU, United Kingdom}
\newcommand{\twaffil}{Physics of Fluids Group, Max Planck Center for Complex Fluid Dynamics, MESA+ Institute and J. M. Burgers Center for Fluid Dynamics, University of Twente, PO Box 217,7500AE Enschede, Netherlands}

\title{Chemotactic self-caging in active emulsions}
\author{Babak Vajdi Hokmabad~\orcid{0000-0001-5075-6357}}%
\affiliation{\goeaffil}
\author{Suropriya Saha~\orcid{0000-0001-6029-4141}}
\affiliation{\goeaffil}
\author{Jaime Agudo-Canalejo~\orcid{0000-0001-9677-6054}}
\affiliation{\goeaffil}
\author{Ramin Golestanian~\orcid{0000-0002-3149-4002}}
 \email{ramin.golestanian@ds.mpg.de}%
\affiliation{\goeaffil}
\affiliation{\oxaffil}
 \author{Corinna C. Maass~\orcid{0000-0001-6287-4107}}%
 \email{corinna.maass@ds.mpg.de}%
\affiliation{\goeaffil}
\affiliation{\twaffil}
\date{\today}%

\begin{abstract}
A common feature of biological self-organization is how active agents communicate with each other or their environment via chemical signaling. 
Such communications, mediated by self-generated chemical gradients, have consequences for both individual motility strategies and collective migration patterns.
Here, in a purely physicochemical system, we use self-propelling droplets as a model for chemically active particles that modify their environment by leaving chemical footprints, which act as chemorepulsive signals to other droplets. 
We analyze this communication mechanism quantitatively both on the scale of individual agent-trail collisions as well as on the collective scale where droplets actively remodel their environment while adapting their dynamics to that evolving chemical landscape. 
We show in experiment and simulation how these interactions cause a transient dynamical arrest in active emulsions where swimmers are caged between each other's trails of secreted chemicals. Our findings provide new insight into the collective dynamics of chemically active particles and yield principles for predicting how negative autochemotaxis shapes their navigation strategy.

\end{abstract}

\maketitle
Motile micro-organisms have evolved to sense their environment and react to  external chemical or physical cues via \textit{taxis}. Specifically, organisms respond to a gradient in the concentration field of a chemical species by chemotaxis~\cite{adler1966_chemotaxis}, or auto-chemotaxis when the gradient is generated by the organisms themselves~\cite{mittal2003_motility}. In microorganisms, chemotaxis and autochemotaxis guide many collective processes such as colony migration~\cite{fu2018_spatial,bhattacharjee2021_chemotactic}, aggregation~\cite{brenner1998_physical,budrene1995_dynamics}, or biofilm formation \cite{gelimson2016_multicellular}, where the emergent complex behavior is governed by the interplay of physical effects and biological processes. 
Many aggregatory, quorum-sensing~\cite{laganenka2016_chemotaxis} behaviors are based on attractive signaling, \textit{i.e.\ }positive autochemotaxis. Repulsive signaling (negative autochemotaxis) is of practical importance to efficient space exploration, \textit{e.g.\ }when ant colonies forage using mutual avoidance~\cite{robinson2005_no}.

Complex collective behavior can result from intricate biological mechanisms, but also be solely caused by non-equilibrium dynamics (see Refs. \cite{Gelimson2015,Mahdisoltani2021,jin2017_chemotaxis,jin2017_chemotaxis,kruger2016_dimensionality} for such examples), such that there is a need to untangle physics and biology.
To this end, current research in artificial active matter aims to design and develop synthetic micro-swimmers that can mimic strategies like chemotaxis by purely physicochemical means \cite{golestanian2019}.   
A popular swimmer paradigm are self-phoretic particles, which propel via a self-generated local chemical gradient~\cite{howse2007_self-motile,illien2017_fuelled,niu2017_self-assembly}. Suspensions of these particles exhibit non-trivial dynamics influenced by autochemotaxis~\cite{agudo-canalejo2019_active,theurkauff2012_dynamic,golestanian2012_collective,pohl2014_dynamic,solovev2013_collective,liebchen2015_clustering}.  
Specifically, self-propelling droplets~\cite{maass2016_swimming} provide an experimental model for repulsive chemical signaling~\cite{jin2017_chemotaxis,jin2019_fine,moerman2017_solute-mediated,meredith2021_chemical}: Along their way, the droplets shed a persistent trail of depleted fuel which acts as a chemorepellent to other droplets. Hence, the motion of such a droplet is affected by the previous passage of another droplet. 

In this study, we show that in an active emulsion, droplets remodel their chemical environment while adapting their dynamics to that evolving resource landscape. 
In spirit, this resembles \textit{P. aeruginosa} organizing their interactions by shedding attractive trails, \cite{kranz2016_effective,gelimson2016_multicellular}, with the difference that droplet trails are chemorepulsive.
    
We start by a quantitative analysis of individual `delayed collision' events: we directly visualize and map the chemical footprints of droplets and measure the diffusion coefficient of the constituent chemicals.  We use these results to fit a generic analytical model based on time delay, angle of impact and chemical coupling strength. We show that these parameters determine whether a droplet crosses a chemical trail or rebounds from it.
We then proceed to the collective dynamics, comparing experimental data to simulations using our single-event fits. We demonstrate how such individual binary collisions cause autochemotactic arrest in ensembles of chemically-active droplets, a kind of `history caging', where swimmers are transiently trapped in an evolving network of repulsive trails. 

\section*{Results}
\subsection*{Self-propulsion by solubilisation}

We study active emulsions containing microdroplets of CB15 oil in an aqueous, strongly supramicellar aqueous solution of the surfactant TTAB. 
The gradual solubilisation of the oil into the surfactant micelles causes droplet activity: While the solubilisation is radially isotropic below a threshold surfactant concentration, above the threshold this symmetry is broken by a spontaneously emerging front-back gradient in the surfactant coverage of the interface, imposing, via the respective interfacial tension gradient (Marangoni effect), an orientation $\bm{n}$ or polarity onto the droplet, and leading to self-propulsion in the direction of $\bm{n}$. 
This can be readily understood if we assume that an increased fraction of oil-filled vs. empty micelles ($\rho_\text{fm}/\rho_\text{em}$) in the vicinity depletes the interface of surfactant~\cite{herminghaus2014_interfacial,maass2016_swimming,meredith2020_predator-prey}. 
Therefore, the trail of oil-filled micelles in the wake of a moving droplet locally increases the interfacial tension (\figref[D]{fig:TrailVis}), providing positive feedback to the front-back tension gradient that leads to self-supporting motion. This sensitivity of the interface to local changes of  $\rho_\text{fm}/\rho_\text{em}$ also applies to the trails left by other droplets: $\bm{n}$ is continuously reoriented towards areas of lower $\rho_\text{fm}/\rho_\text{em}$~\cite{jin2017_chemotaxis,moerman2017_solute-mediated}. 

\subsection*{Quantification of the chemical trail behind the droplet}

\begin{figure*}
	\includegraphics[width=0.9\textwidth]{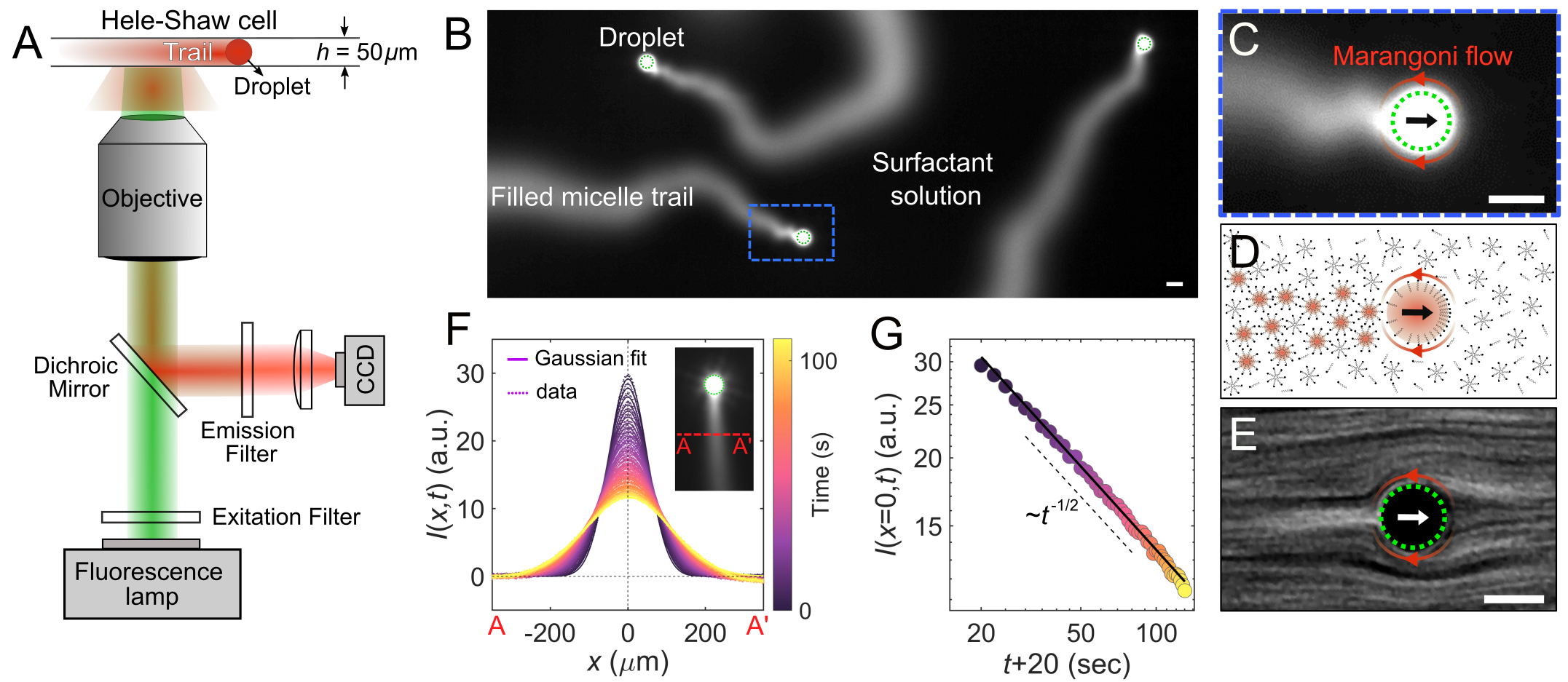}
    \caption{\textbf{Visualization of the chemical trail.} \plb{a} Schematic of the experimental setup for fluorescent microscopy of the filled micelle trail. CB15 droplets of diameter $a_\text{drop}=\SI{50}{\um}$ were injected into a quasi 2D microfluidic cell (height= \SI{50}{\um}) and observed using either bright field or fluorescent microscopy. \plb{b} A fluorescence micrograph of the droplet's chemical trails, with surfactant concentration increased to 15\,wt.\% to improve trail visibility (increased solubilisation rate). \plb{c} Zoomed-in view of \plb{b}. 
    \plb{d} Schematic propulsion mechanism of the droplet. The black arrow shows the direction of motion. \plb{e} The flow field generated by Marangoni flow at the droplet interface visualized by streak lines of 0.5 $\mu \rm m$ fluorescent tracer colloids (droplet reference frame).
    \plb{f} The time evolution of fluorescent intensity profiles along $\textrm{AA}'$ (see inset) superimposed with Gaussian fits (surfactant concentration 5\,wt.\%). \plb{g} Peak intensity versus time. The zero point in time is shifted by 20 seconds from the droplet passage time to account for the fact that the droplet is not a point source $ I_0\delta(x)$  (see Methods). All scale bars are \SI{50}{\um}, green circles mark the droplet boundary in overexposed areas. }
	\label{fig:TrailVis}
	\end{figure*}

To measure this chemical trail we added the hydrophobic fluorescent dye Nile Red to the oil phase~\cite{hokmabad2021_emergence} of a droplet swimming inside a quasi-2D microfluidic cell. The dye molecules co-migrate with the oil into the swollen micelles left in the wake of the droplet. We recorded and quantified the distribution of oil-filled micelles via fluorescent microscopy (see \figref{fig:TrailVis} and video S1).

To model the trail diffusion, we approximate the droplet as a point source moving at speed $V_0\approx\SI{26\pm5}{\um\per\second}$ and emitting a fluorescent chemorepellent at a constant rate. Accordingly, the fluorescence intensity along a fixed line perpendicular to the trail (e.g. $\textrm{AA}'$ in \figref[F,inset]{fig:TrailVis}) should be Gaussian with a peak height scaling with  $t^{-1/2}$. 
The measured time-dependent intensity along $\textrm{AA}'$  in \figref[F,G]{fig:TrailVis} shows excellent agreement with the Gaussian model, yielding a diffusion constant for filled micelles $D_\text{fm}=\SI{52.5}{\um^2\per\second}$ (details in SI). This is further consistent with the Stokes-Einstein relation $D_\text{fm}={k_BT}/{(6\pi \eta r_\text{fm})}=\SI{55.2}{\um^2\per\second}$ for $r_\text{fm}=\SI{3}{\nano\metre}$ and literature values~\cite{jin2017_chemotaxis,izzet2020_tunable}.
We estimate the diffusive time scale of the trail spreading as $\tau_\text{diff}=a_\text{drop}^2/D_\text{fm}\approx\SI{45}{\second}$, which is considerably longer than the droplet's advective time scale $a_\text{drop}/V_0\approx\SI{2}{\second}$. Such long-lived chemical gradients influence the propulsion dynamics of other droplets even after hydrodynamic interactions have decayed. Beyond the advective timescale, we can therefore model droplet-trail interactions by a `dry' chemically-active polar particle (CAPP) approach~\cite{saha2014_clusters,saha2019_pairing}, as follows.

\subsection*{Chemically Active Polar Particle model}

The fore-aft flow at the interface is determined by the gradient in the local concentration of the empty micelles. We assume that, in the steady swimming state, the fore-aft asymmetry of the droplet is stably maintained. Thus, when a droplet encounters a trail, it experiences an effective torque, governed by a coupling constant $\Omega$, and an effective force, with coupling constant $\alpha$.

Denoting the position of droplet $i$ by $\bm{r}_i$ and its orientation by $\bm{n}_i$, their Langevin dynamics are as follows \cite{saha2014_clusters}:

\begin{align}
\dot{\bm{r}}_i &= V_0 \bm{n}_i - \alpha \bm{\nabla} c |_{\bm{r}=\bm{r}_i} + \sqrt{2 D_t} \,\bm{\xi}_t, \label{eq:posDynamics} \\
\dot{\bm{n}}_i &= \Omega \,\bm{n}_i \times (\bm{n}_i \times \bm{\nabla} c|_{\bm{r}=\bm{r}_i}) + \sqrt{2 D_r}\, \bm{n}_i \times \bm{\xi}_r,
\label{eq:axisDynamics}
\end{align}
where $c(\bm{r},t)$ is the micelle concentration field, $\bm{\xi}_t$ and $\bm{\xi}_r$ represent Gaussian distributed translational and rotational noise with unit strength, and $D_t$ and $D_r$ are the corresponding translational and rotational diffusion coefficients of the droplet. The analysis of the mean squared displacement of the droplets in dilute conditions reveals ballistic trajectories over periods larger than \SI{100}{\s} (see ~\figref[B]{fig:2DCaging}), suggesting that Brownian translational diffusion is negligible, and that the rotational diffusion coefficient is also small, with an upper bound of $D_r = \SI{0.01}{\radian^2\per\s}$ corresponding to persistence lengths larger than $V_0/D_r\approx \SI{2.5}{\milli\metre}$. We note that the general theory of CAPPs also allows for an effective force projected along the axis of the particle \cite{saha2014_clusters}, with a third coupling constant, but we will neglect this possibility here as the data can be explained within the more minimal model with two coupling constants.

To model the micelle concentration field $c(\bm{r},t)$, we use a simplified static one-dimensional Gaussian profile perpendicular to the direction of motion when studying individual droplet-trail interactions (see Appendix, \ref{SIsec:modelsingle}), and a full description of the dynamic two-dimensional concentration field when modeling the collective behavior (see Appendix, \ref{SIsec:modelmany}).

\subsection*{Quantitative analysis of droplet-trail interactions}

We now proceed to study the individual droplet-trail interactions in a quasi-2D microfluidic geometry (details in Appendix,~\ref{SIsec:improcessing}). 
\figref[A,B]{fig:Autochemotactic} show examples of crossing and reflecting events, respectively (videos S2 and S3). 
The following droplet (red trajectory) approaches at an incident angle $\theta_{\rm inc}$ with respect to the first trajectory (blue).
An interaction starts and ends when the distance $|d|$ between the droplet and the trail falls below a threshold value $d_\text{max}=\SI{220}{\um}$.
We identify the points of intersection (green points), or, for reflection, closest approach on each trajectory, and define the time lag $\Delta t$ as the interval between each droplet passing these points, and the time origin $t_0$ as the respective point in time for the following droplet. We observe, for similar $\theta_\text{inc}$, a transition from reflection to crossing with increasing $\Delta t$.

\begin{figure*}
	\centering
	\includegraphics[width=0.9\textwidth]{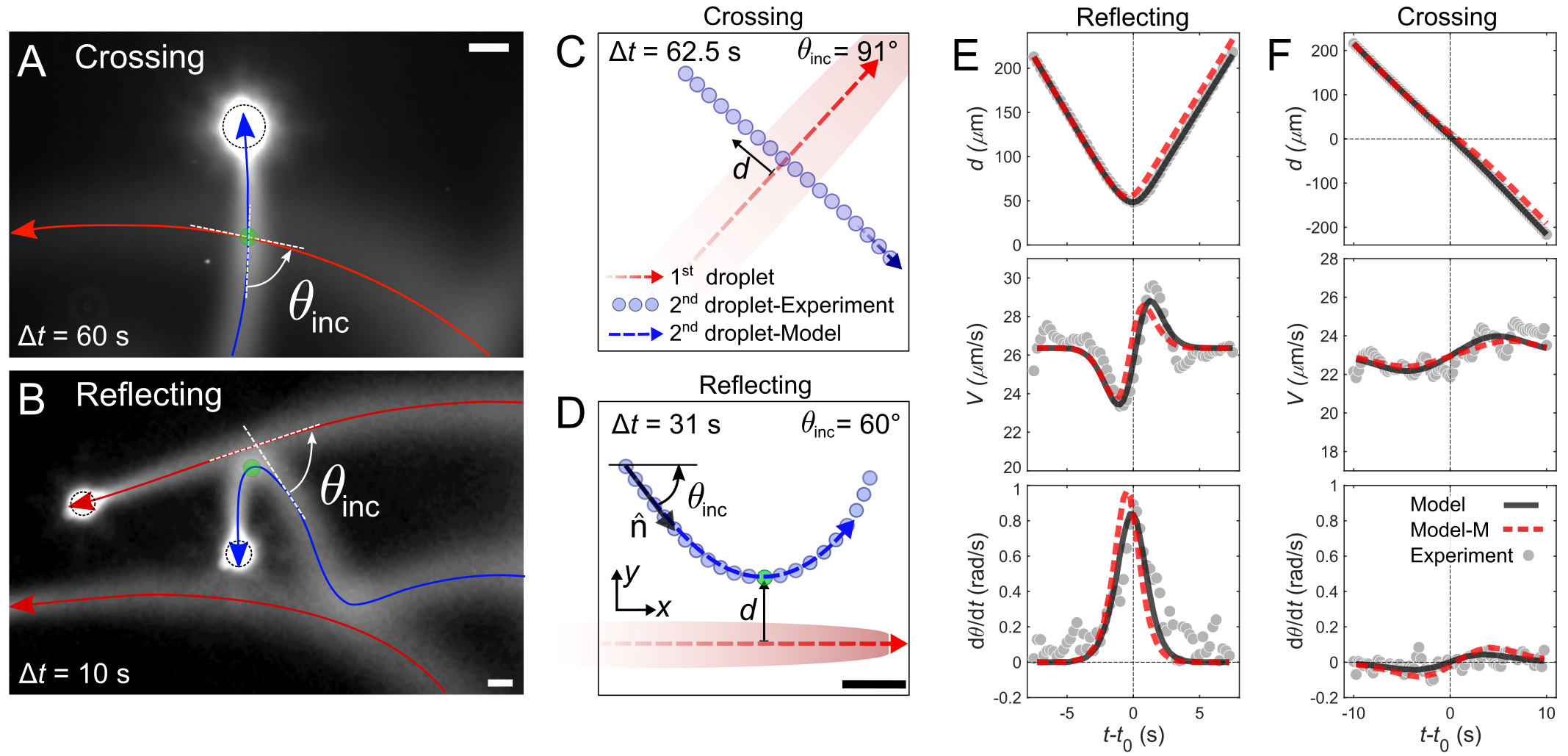}
    \caption{\textbf{Autochemotactic interaction between a droplet and a trail.}  Fluorescent micrographs of a crossing \plb{a} and a reflecting \plb{b} interaction. The red trajectory corresponds to the first passing droplet secreting the trail, the blue one to the following droplet.
    \plb{c} and \plb{d} Trajectories from bright field microscopy for one crossing and one reflecting interaction. Dashed lines are the theoretical fits from the CAPP model, using fit parameters $\Omega c_0$ and $\alpha c_0$.  \plb{e} and \plb{f} Plots of signed distance $d$, swimming speed $V$ and rotation rate $\rm{d}\theta/\rm{d}t$ (angular velocity) for typical reflecting \plb{e} and crossing \plb{f} interactions, respectively. 
    Black lines (Model) correspond to the best fit for the given trajectory, red dashed lines (Model-M) to the theoretically predicted trajectories using the median values of all fits analyzed, $\Omega c_0 = 7 \cdot 10^3$ $\mu$m$^2/$s and $\alpha c_0 = 3 \cdot 10^4$ $\mu$m$^3/$s.
    All scale bars are \SI{50}{\um}.}
	\label{fig:Autochemotactic}
\end{figure*}

To match the CAPP model to the experimental data we analyzed ``delayed collisions'', as shown in the examples in \figref[C,D]{fig:Autochemotactic}. For the following droplet, we measured the signed distance $d$ from the trail, speed $V$ and angular velocity $|\rm d\theta/d\rm t|$ (\figref[E,F]{fig:Autochemotactic}). We fit the theoretical trajectories to the experimental trajectories in order to estimate the two unknown coupling constants $\Omega c_0$ and $\alpha c_0$, which were used as fitting parameters (see Appendix,~\ref{SIsec:modelsingle} for details).  Examples of the fits are shown in \figref[E,F]{fig:Autochemotactic} for two experimental trajectories. Here, black lines correspond to the actual fit to the given trajectory, and red lines correspond to the theoretical prediction using the optimal median values obtained from all fits in the data set, $\Omega c_0 = \SI{7e3}{\um^2\per\second}$ and $\alpha c_0 = \SI{3e4}{\um^3\per\second}$. The characteristic features of the evolution of the position, velocity, and angular velocity with time are well recapitulated by the model, both for each particular fit and when using the median values. Importantly, $\Omega>0$ implies that the droplet reorients to point away from the trail, whereas $\alpha > 0$ implies that the droplet is also directly repelled by the trail. While we found that the trajectory shapes are most sensitive to changes in $\Omega$, which is the key parameter governing the interaction, the presence of a positive $\alpha$ is essential in order to correctly capture the time evolution of the droplet velocity (second row in \figref[E]{fig:Autochemotactic}), which decreases before the turning point and increases after it.

\subsection*{Incidence angle and time lag determine the interaction dynamics}

\begin{figure}
\centering
	\includegraphics[width=\linewidth]{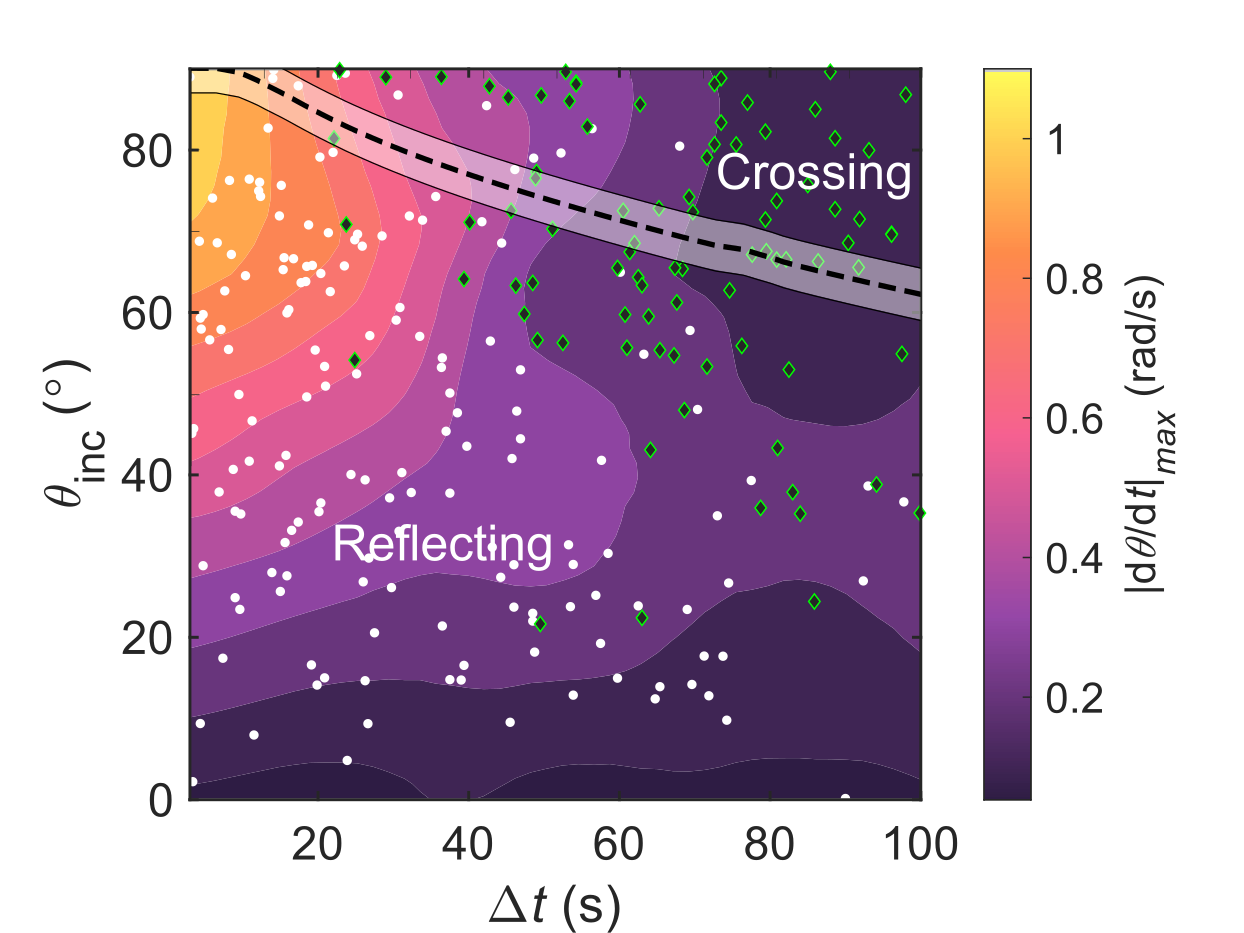}
		\caption{\textbf{Phase diagram of autochemotactic interactions}, mapping 164 trail reflections (white) and 90 crossings (green). The interpolated background color corresponds to the maximum rotation rate $|\rm d\theta/d\rm t|_{\rm max}$ measured for each interaction. The dashed line is the separatrix between crossing and reflection as calculated from fits to the CAPP model (with error interval accounting for rotational diffusion). \label{fig:phasediagrams}}
	\end{figure}
	
Investigating the autochemotactic interactions revealed that the probability of crossing versus reflection depends both on $\theta_\text{inc}$ and on $\Delta t$, which determines the gradient strength.
We extracted $\theta_\text{inc}$ and $\Delta t$ for all 254 identified interactions and plotted them in the phase diagram in \figref{fig:phasediagrams}, with reflections marked by white and crossings by green data points. The background color map interpolates the maximum recorded angular velocity.
We can confirm two key features: For one, crossing is more likely for increasing $\Delta t$, i.e. decreasing $\nabla c$. Second, for increasing $\theta_\text{inc}$, i.e. sharper reorientations, reflection requires a higher turning rate, mediated by $\Omega|\nabla c|$, such that crossing is more likely for increasing $\theta_\text{inc}$.

We now construct a theoretical phase diagram using the median values of $\Omega c_0$ and $\alpha c_0$. The separatrix (dashed black line) between crossing and reflecting events, which coincides with those trajectories for which the turning point is exactly at the center of the trail ($y=0$), reproduces the salient features of the experimental observations. The small amount of rotational noise present in the system can cause some uncertainty in the location of the separatrix, which we estimate from the standard deviation of the orientation while the incoming droplet approaches the trail: $\delta \theta \approx \sqrt{2 D_r T}$, using an approach time $T \approx d_\mathrm{max}/(V_0 \sin \theta_\mathrm{inc})$ and an upper bound of $D_r \approx 0.01$ rad$^2$/s for the rotational diffusion coefficient.  The strongest deviations between the theoretical prediction and experimental results are observed at low $\theta_\mathrm{inc}$ and large $\Delta t$, in which case both crossing and reflecting trajectories are seen to coexist experimentally. We expect that in this region higher order effects, such as the time-dependence and space-dependence of the trail width (which breaks the symmetry between parallel and antiparallel trajectories) will play a role.

\subsection*{Collective dynamics governed by autochemotactic interactions: history caging}

To study the consequences of autochemotactic interactions in a crowded system, we placed suspensions of swimmers at number densities $n$ between $0.025$ and \SI{8.6}{\per\milli\metre\squared} in a quasi-2D cell and recorded their trajectories for longer times $(\approx \SI{5}{\minute})$.  

We first illustrate the collective behavior using a fluorescently dyed sample in \figref[A]{fig:2DCaging} (video S4), at increased solubility for better visualization.   
Initially, all droplets move persistently, but reorient when they encounter a trail. Gradually, the secreted trails form a network. A complex chemical landscape, based on $\nabla c$, evolves in time and space with multiple local minima (dark regions) between the trails.
The swimmers get transiently trapped in these interstitial minima by multiple reflections at the cage walls (video S5). They can escape either when $\nabla c$ has decreased sufficiently at the cage boundaries or when the chemical buildup caused by the droplet itself forces it out of the cage (video S6).

We now quantify the collective dynamics under the same chemical conditions as the droplet-trail interaction experiments.
\figref[B]{fig:2DCaging} shows, for increasing number densities, the mean squared displacement obtained by ensemble averaging over the trajectories, 
\begin{align}
\text{MSD}(t) = \frac1N\sum_{i=0}^{N}\left(\vec{r}_i(t)-\vec{r}_i(t_0)\right)^2,
\label{eq:MSD} 
\end{align}

where $t_0$ is the starting time of the experiment and $N$ the number of droplets. 
For any number density, droplets initially move persistently, with $\text{MSD} \sim t^2$. For an isolated droplet, we do not see a transition to diffusive scaling $\sim t$ on our experimentally accessible length scales. 
For intermediate number densities, we observe a change in the slope of the MSD, which is associated with the reorientations caused by the autochemotactic interactions. 
At large number densities ($ n\geq3.7 $ \SI{}{\per\milli\metre\squared}), after a short ballistic period, the MSD reaches a plateau which is a signature of caging. Such a plateau is reminiscent of the caging phenomenon in colloidal glasses~\cite{weeks2000_three-dimensional}. However, here the caging is caused by trail-droplet interactions instead of direct interparticle collisions and is therefore observed at much lower volume fractions ($\phi_\text{droplets}\approx 10^{-2}$, cf. \cite{pusey1986_phase}).  For more crowded systems, the crossover to caging happens earlier, the lifetime of the cage is longer, and the cage size is smaller.  The plateau is followed by a crossover to a third, subballistic regime corresponding to the droplet escaping the cage. We demonstrate this for an example trajectory in \figref[C]{fig:2DCaging} (thick markers), where a droplet undergoes three caging events (at $ n\approx \SI{7.4}{\per\milli\metre\squared}$). To highlight the cage formation by spatiotemporal aggregation, we plotted all droplet positions (thin symbols) recorded within a window of  $d<\SI{220}{\micro\metre}$ and $0<\Delta t<\SI{50}{\second}$ around each current droplet position. Entering an area with increased density, the droplet reorients frequently, exploring the cage, until it is ejected due to the gradual buildup of chemorepellent into a less populated space (`cage escape'), where it proceeds persistently until it encounters the next high-density area (cf. Movie S6).

We recover this caging behavior in numerical simulations of the CAPP model (see Appendix~\ref{SIsec:modelmany} for details) using only the experimentally measured parameters as well as the median $\Omega c_0$ and $\alpha c_0$ extracted from the fits to individual particle-trail interactions, without any adjustable parameters (\figref[D,E]{fig:2DCaging}). We observe the emergence of a subdiffusive `caging' regime at the same particle densities and with comparable caging timescale and length scale as in experiments. The simulations further confirm that beyond the caging regime, the MSD transitions to a diffusive regime as expected.

\begin{figure*}
	\includegraphics[width=0.95\linewidth]{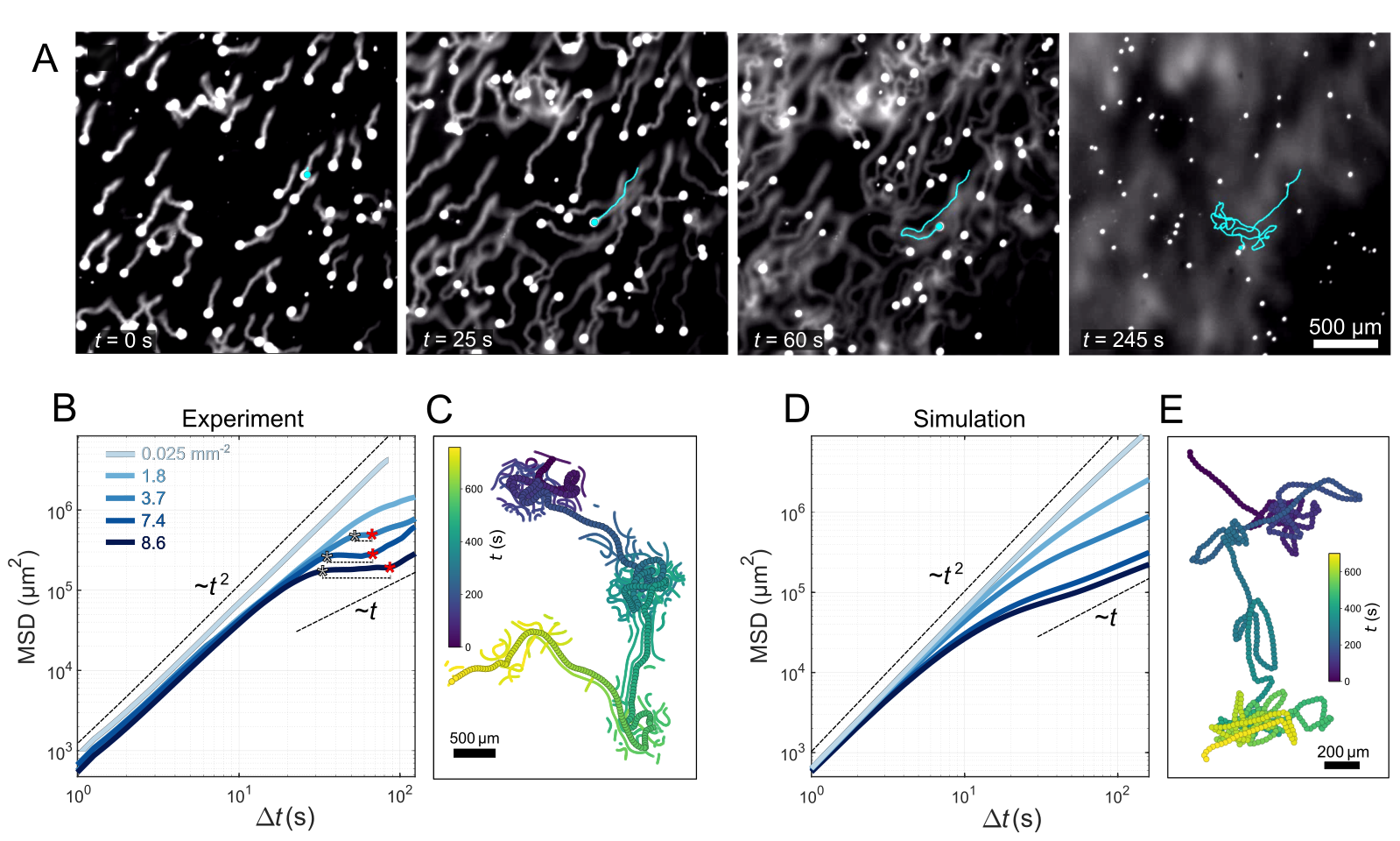}
	\caption{\textbf{Caging in 2D.} \plb{a} Snapshots from fluorescent microscopy of an active emulsion in a quasi-2D cell (droplet size $\SI{50}{\micro\metre}$), under high droplet solubilisation rate to increase trail fluorescence. The cyan trajectory demonstrates the evolution from ballistic propulsion to caging. 
	\plb{b} Mean squared displacements for emulsions with increasing number densities. The $\star$ denotes the cross-over to the caging regime, \color{red}$\star$ \color{black} the cross-over to the cage-escape regime. \plb{c} A trajectory $s(t)$ with consecutive caging and cage-escape events (thick symbols). Thin symbols: other droplets detected within a spatiotemporal window of $d<\SI{220}{\micro\metre}$ and $0<t_0-t<\SI{50}{\second}$ around the current trajectory point $s(t_0)$. Data color coded by time. \plb{d} MSD and \plb{e} example trajectory obtained from simulations of the CAPP model under the same conditions as the experiments in (b,c), demonstrating similar caging behavior.}	\label{fig:2DCaging}
\end{figure*}

\subsection*{Caging in 3D}
An intriguing question is how far dimensionality matters in the emergence of caging. In two-dimensional confinement, fully confining cages exist even for point-like particles and one-dimensionally parameterized trajectories. In three dimensions, such a cage would always have holes the droplet can escape through. Caging in 3D is therefore only possible owing to the finite volume of the diffusing trail.

To investigate caging in 3D, we placed active emulsions of varying number densities $n$ in a swimming medium density matched by heavy water admixture. We recorded the 3D trajectories of droplets inside a (3mm)$^3$ volume using a scanning light sheet fluorescence microscope, as shown in  
\figref[A]{fig:3DCaging} by 3D rendering of trajectories recorded over $\approx\SI{6.3}{\minute}$ and the droplet arrangement after \SI{29}{\second} at  $n=\SI{8}{\per\cubic\milli\metre}$.
We have plotted typical trajectories for systems with increasing $n$ in \figref[B-D]{fig:3DCaging}. In the dilute case, $n=\SI{2}{\per\cubic\milli\metre}$, the trajectory is quite straight, while, at intermediate $n=\SI{2}{\per\cubic\milli\metre}$, the droplet undergoes a few reorientation events. The trajectory for the densest system at $n=\SI{22}{\per\cubic\milli\metre}$ shows alternating straight and caged sections similar to the behavior in 2D (\figref[D]{fig:2DCaging}).

In \figref[E]{fig:3DCaging}, we plotted the mean squared displacement of a set of 3D trajectories extracted from the data sets used for \figref[B-D]{fig:3DCaging}. The signatures of caging can be observed in the form of plateaus in the MSD, in particular for $n=\SI{22}{\per\cubic\milli\metre}$.

\begin{figure*}[ht]
     \centering
	\includegraphics[width=.6\textwidth]{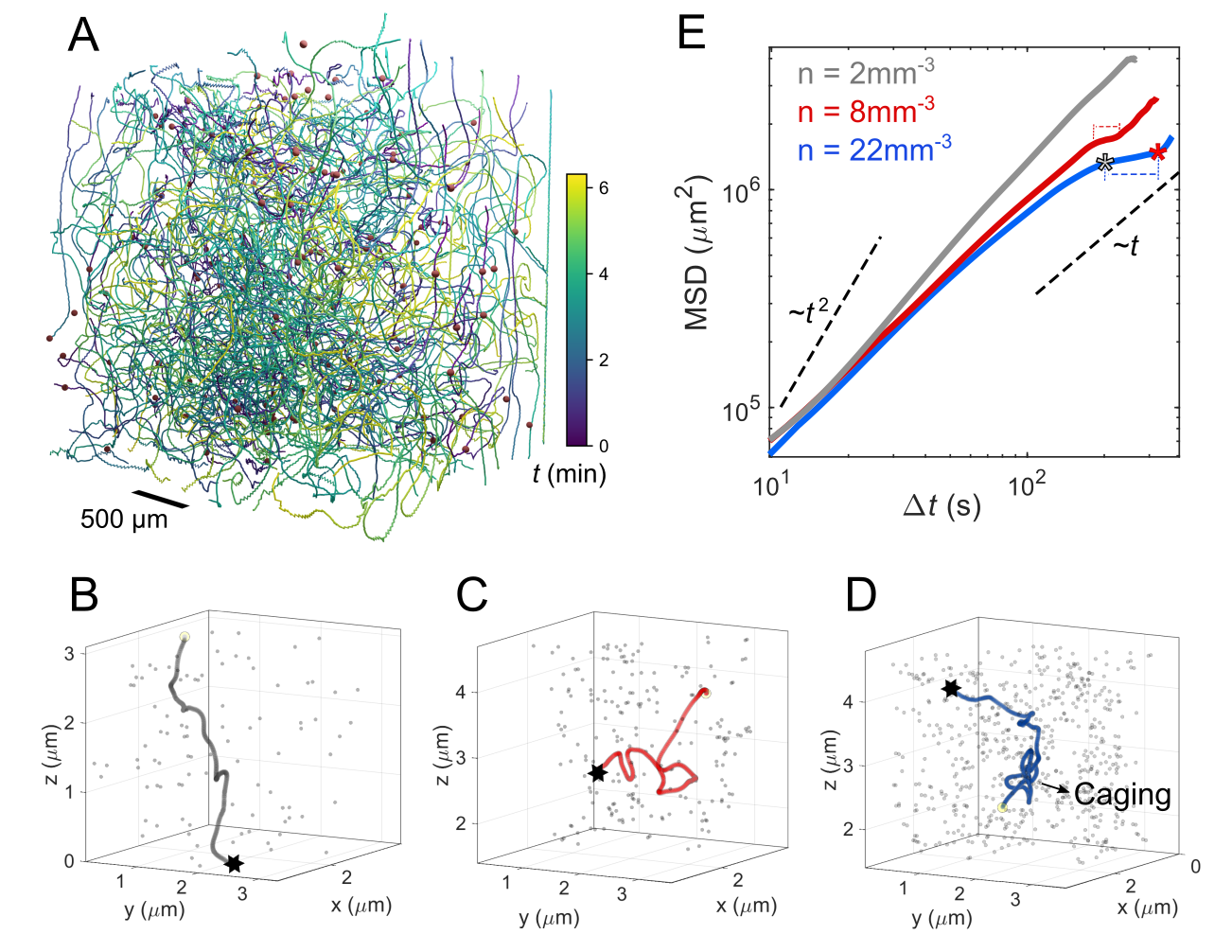}
	\caption{\textbf{Caging in 3D.}  \plb{a} Tracking by fluorescent light sheet microscopy. 3D reconstruction of trajectories recorded over $\approx\SI{6.3}{\minute}$ and the droplet arrangement (red spheres) after \SI{29}{\second} at number densities $n=\SI{8}{\per\cubic\milli\metre}$. Trajectories color coded by time. Sample volume (3mm)$^3$, swimming medium and droplet density matched by $\rm D_2\rm O$ admixture to enable force-free swimming.  \plb{b-d} 3D trajectories for one droplet swimming in emulsions with increasing number density 2, 8, and \SI{22}{\per\cubic\milli\metre} recorded during 221 s, 380 s, and 490 s, respectively . End points are marked by $\star$. Grey markers show one typical arrangement of all droplets.  \plb{e} Mean squared displacements of 3D active emulsions with increasing $n$.} 	\label{fig:3DCaging}
\end{figure*}

\section*{Discussion}

We have used active emulsions to explore the collective behavior of active particles governed by negative autochemotaxis. Using fluorescent imaging, we visualized and measured the diffusion coefficient of the chemorepellent trails left in the wake of the droplets; a quantity necessary for theoretical modeling of the interactions. We quantified these interactions and established a generic theoretical CAPP model that can recreate the trajectories and predict the outcome of a droplet-trail collision based on three parameters: time lag between droplet passages, angle of incidence, and chemical coupling strength. We would like to highlight the fact that the emergent behavior observed experimentally at the collective level is recovered in the theoretical model without the need for any fitting, as discussed above. This suggests that the basic ingredients of the model---and in particular the alignment interaction (torque) term with a strong value for its coupling $\Omega$---are presenting a faithful minimal theoretical description of the system, with predictive power at the collective level. 

While it is in principle feasible to treat individual droplet-trail interactions including hydrodynamic feedback~\cite{lippera2020_alignment,lippera2020_collisions}, our results show that the CAPP model captures essential features of the delayed collisions in dilute systems, where, typically, the time lag between droplet passages is considerably longer than the advective timescale. As shown here, this makes the collective behavior accessible to numerical and analytical modeling using a similar simplified paradigm. 

We then studied the collective dynamics in active emulsions and observed a dynamical arrest mechanism: autochemotactic caging in both quasi-2D and unconfined 3D geometries. 
Our observations show that, even in 3D, droplets can get trapped in an evolving chemical landscape created by the trails of other swimmers. Their dynamics resemble the recently reported bacterial hopping and trapping in a heterogeneous  porous medium~\cite{bhattacharjee2019_bacterial}, however, in this case the heterogeneous medium is self-generated. Remarkably, we see an enhancement of density fluctuations reminiscent of quorum sensing~\cite{bauerle2018_self-organization}, even based on purely repulsive interactions.
Compared to 2D confinement, 3D caging sets in at significantly lower volume fractions ($\phi\approx 10^{-4}$).
Here, we hypothesize that the droplet motion in 3D is not rectified by the cell boundaries and that reflection is therefore effected by weaker gradients. An in-depth answer to this question will require further quantitative modeling of the individual interactions in 3D.

Due to the generality of the theoretical framework, the current approach and the insights obtained from this study can be extended to similar systems  where active agents locally deplete the resources and in turn remodel their environment. The interactions of the agents with the evolving resource landscape~\cite{thakur2011_dynamics} result in the emergence of complex collective states such as the caging reported here, adaptive search strategies~\cite{volpe2017_topography}, or field-driven localization for ecology-inspired robots~\cite{wang2021_emergent}.

\section*{Acknowledgments}

We acknowledge fruitful discussions with Stephan Herminghaus and Carsten Krüger. C.C.M. and B.V.H. acknowledge financial support by the DFG SPP1726, `Microswimmers'. R.G. acknowledges support from the Max Planck School Matter to Life and the MaxSynBio Consortium which are jointly funded by the Federal Ministry of Education and Research (BMBF) of Germany and the Max Planck Society.

\appendix

\section{Materials and Methods}

\subsection{Materials and characterization}
We study active oil-in-water emulsions, whose oil phase consists of  CB-15 (Synthon Chemicals), an isotropic isomer of the common nematogenic oil 5CB.  Mono-disperse droplets of diameter $d= \SI{50\pm 3}{\um}$ are mass produced in microfluidic flow focusing devices (see~\cite{hokmabad2021_emergence} for the detailed protocol). For the purpose of chemical quantification, we dissolved small amounts of the fluorescent dye Nile Red (Thermo Fisher Scientific) in the oil phase. Nile Red does not fluoresce in water, such that we can assume the fluorescent intensity to stem from micelles filled with oil and co-migrating Nile Red. This assumption is supported by the diffusive spread of fluorescence in the trail mapped in \figref{SIfig:fluorescence}.
The swimming medium for all quantitative experiments was a 5wt\% aqueous solution of TTAB surfactant (Sigma Aldrich). For force free bulk measurements, the density of 5CB ($\rho_\text{5CB}=\SI{1.05}{\gram\per\cubic\centi\metre}$) was matched by appropriate heavy water substitution ($\rho_\text{D2O}=\SI{1.2}{\gram\per\cubic\centi\metre}$).

\subsection{Methods: observation cell, microscopy and image recording and analysis}

Unless noted otherwise, all experiments were done in microfluidic cells using a quasi 2D Hele-Shaw geometry. We fabricated the cells directly from SU-8 photoresist (Micro Resist Technology) on a glass microscope slide of area $75\times50$ mm$^2$. Rectangular cells with an area of $8\times13\,$mm and height of $50 \mu$m were filled with a suspension of droplets and sealed with a glass cover slip. No air was entrapped in the cell volume, such that we can assume no-slip conditions at all cell boundaries. To study the droplet-trail interactions and the collective behavior we observed active emulsions under a bright-field
microscope (Leica DM4000 B) at a magnification of 2.5x. Videos were recorded at 4 frames per second using a Canon digital camera (EOS 600d) with a digital resolution of 1920 x 1080 px. The droplet coordinates in each frame
were extracted by custom Python scripts (available on request) using the libraries numpy, PIL, and openCV. We obtained droplet coordinates via a sequence of background correction, binarization, blob detection by contour analysis, and minimum enclosing circle fits. Droplet trajectories were obtained using a frame-by-frame nearest-neighbor analysis. 

To directly visualize the filled micelle trail left in the wake of the droplet, we doped the oil phase NileRed and performed fluorescent microscopy on an Olympus IX73 device with a filter cube (excitation filter ET560/40x, beam splitter 585 LP and emissions filter ET630/75m, all Chroma Technology). Images were recorded using a 4 MP CMOS camera (FLIR Grasshopper 3, GS3-U3-41C6M-C) at 4 frames per second and $4\times$ magnification. Further image analysis to extract intensity profiles and quantify the diffusion coefficient were carried out in MATLAB. 
We note that, for illustration purposes, the micrographs in \figref[b,c]{fig:TrailVis},  \figref[a]{fig:2DCaging} and Video S4 were recorded at a higher surfactant concentration of 15 wt.\% to increase the solubilization rate and thereby the trail fluorescence. 
However, a lower concentration (5 wt.\%) leads to more persistent droplet motion and long cruising ranges and is therefore preferable in quantitative experiments. 

\subsection{Micellar diffusion quantification by fluorescence}
\begin{figure}
    \centering
    \includegraphics[width=.33\textwidth]{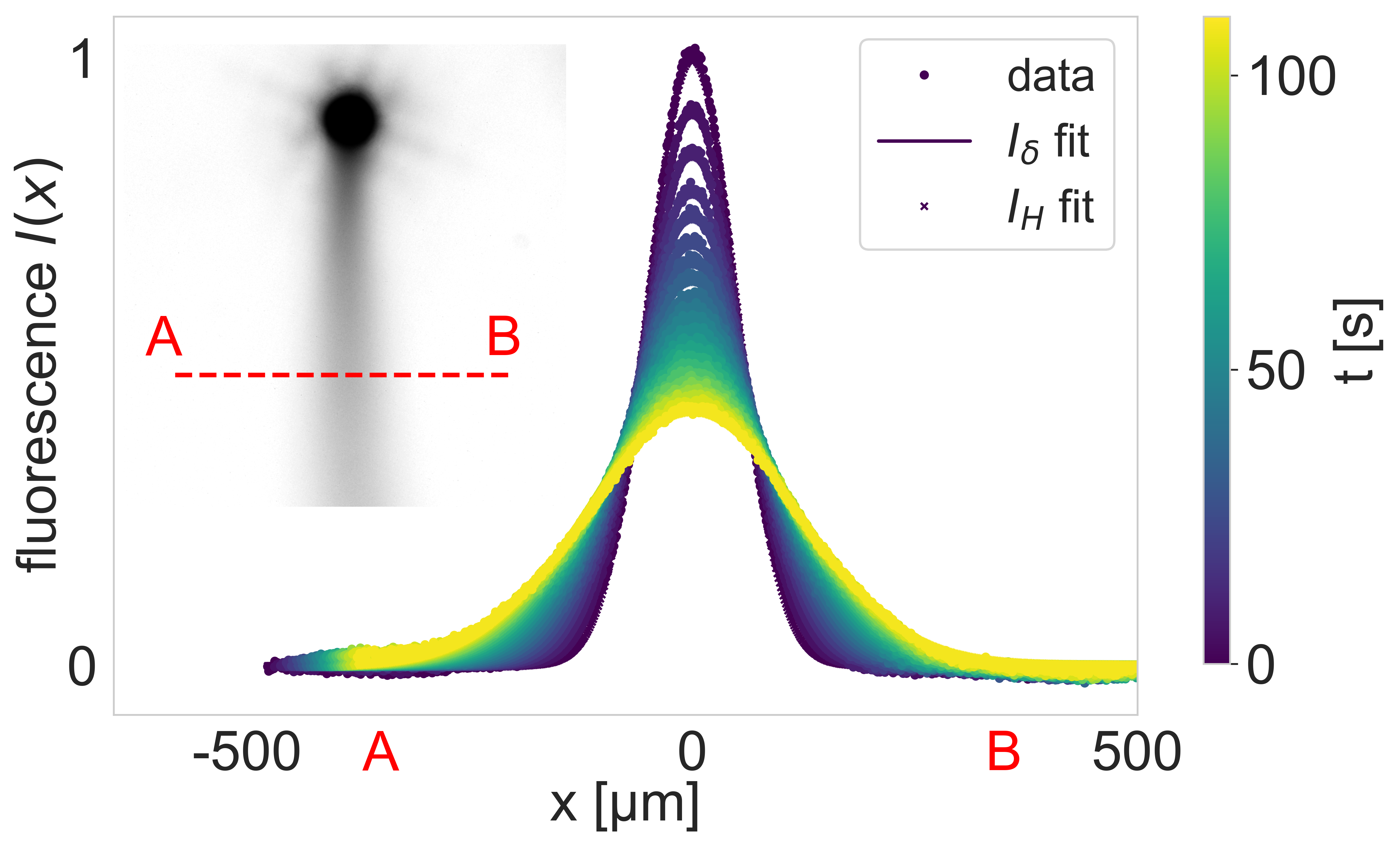}
    \includegraphics[width=.33\textwidth]{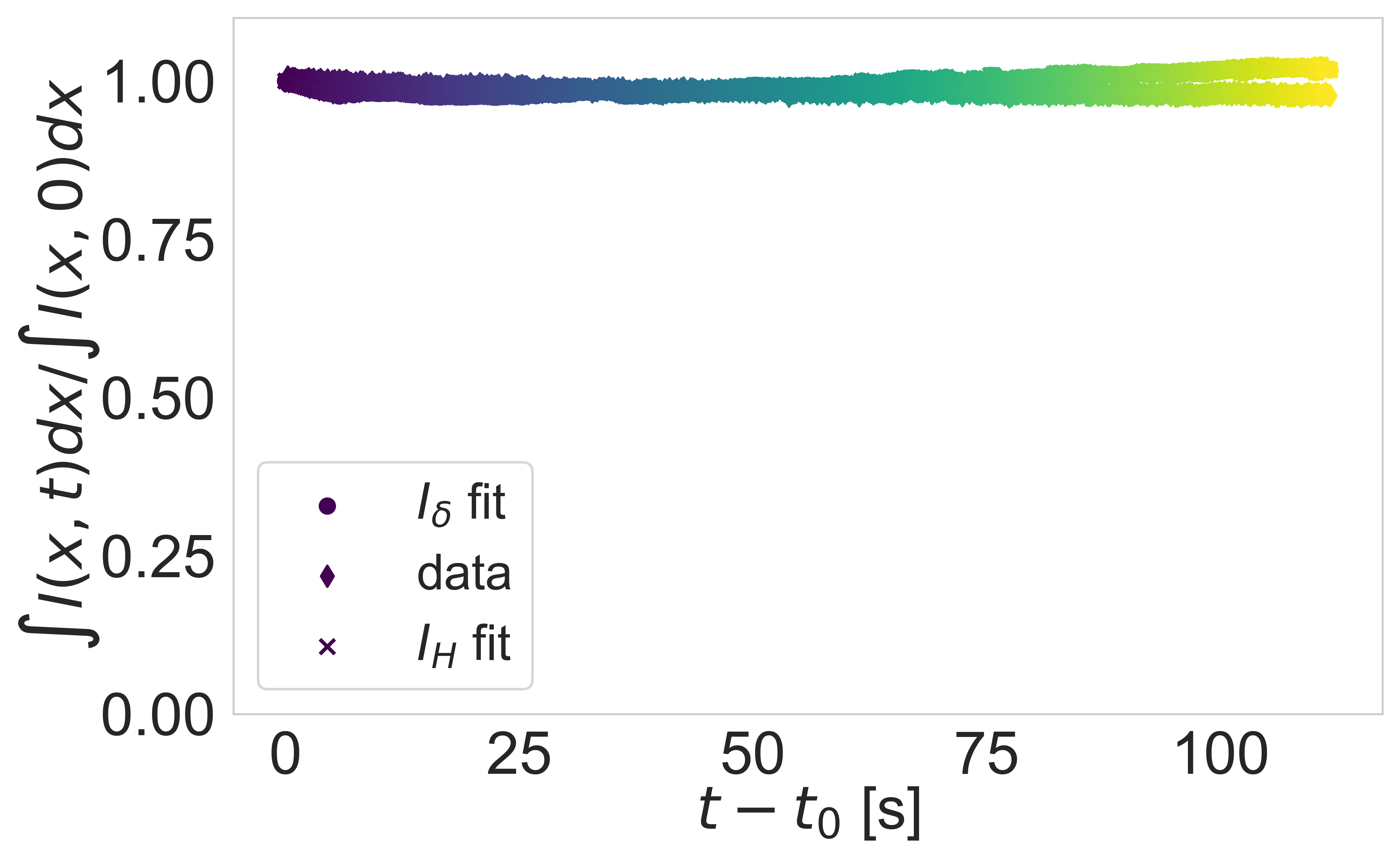}
    \setlength{\unitlength}{.1\textwidth}%
    \begin{picture}(10,0)
    \put(0,4.1) {\small\textbf{(a)}}
    \put(0,2.1) {\small\textbf{(b)}}
    \end{picture}
    \caption{Fluorescent emission in the trail of a Nile Red doped droplet. (a) Experimental data vs. point source  $I_\delta(x,t)$ (\eqref{eq:gauss}) and stepwise $I_H(x,t)$ (\eqref{eq:erf}) emitter fit models. Within experimental precision, the models are indistinguishable. (b) The integrated fluorescence does not decay over time, as shown by numerical integration of the recorded intensity and both fit models. 
    }
    \label{SIfig:fluorescence}
\end{figure}
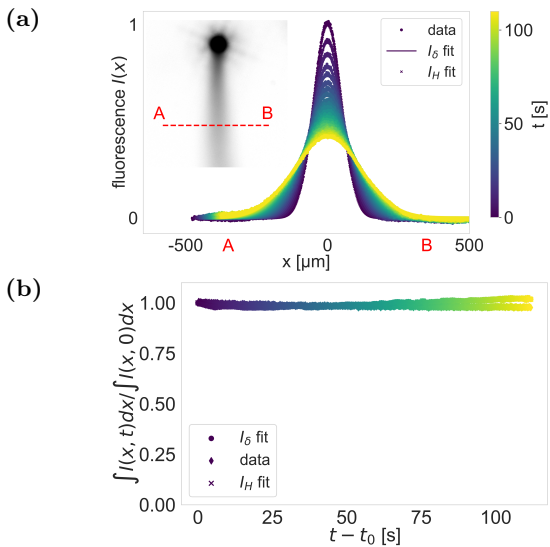

We recorded fluorescent emission from the diffusing trail as follows.
After rotation and rectification of the extracted image series, we extracted profiles along a $y=\text{const.}$ image slice (line A-B in \figref[a]{SIfig:fluorescence}).  We fit the profiles to the time and space dependent decay from a fluorescence quantity $c(x,t=t_0)=M_0\delta(0)$. $t_0$ is earlier than the actual moment of droplet passage, since the droplet is not a point source: 

\begin{align}
        I_\delta(x,t) &= \frac{M_0}{\sqrt{4\pi D_\text{fm}(t-t_0)}}\exp\left(\frac{-x^2}{4 D_\text{fm} (t-t_0)}\right), \label{eq:gauss}\\ M_0&=\int_{-\infty}^\infty I(x,0)\,{\rm d}x
\end{align}

We calculated the micellar diffusion coefficient via modeling the trail as a linear chemical source that was released and started diffusing $t_0$ seconds (obtained from fitting) before the passage of the droplet. We then fit the peak values over time to

\begin{equation}
        I_\delta(0,t) = \frac{M_0}{\sqrt{4\pi D_\text{fm}(t-t_0)}}, \label{eq:gauss_x=0}
\end{equation}

where $x=0$ is the center line of the trail.

To cross-check, we also calculated the micellar diffusion coefficient via $\sigma^2=4D_\text{fm} (t-t_0)$.
In principle, a more precise model could be based on a decay from two Heaviside step functions at a distance $d=\SI{50}{\micro\metre}$, with $c(x,t=0)=c_0\left[H(x+d/2)-H(x-d/2)\right]$:

\begin{equation}
I_H(x,t) = \frac{I_0}{2}\left[\erf\left(\frac{x+d/2}{\sqrt{4 D_\text{fm} t}}\right)-\erf\left(\frac{x-d/2}{\sqrt{4 D_\text{fm} t}}\right)\right]\label{eq:erf}\end{equation}
However, beyond the advective time scale the difference between the models is negligible (\figref{SIfig:fluorescence}).

Since the integrated fluorescence does not decay over time in \figref[b]{SIfig:fluorescence}, we also do not need to account for possible bleaching effects in our analysis.

\subsection{Binary collisions: Image processing and data analysis}\label{SIsec:improcessing}
\begin{figure}
    \centering
    \includegraphics[width=\columnwidth]{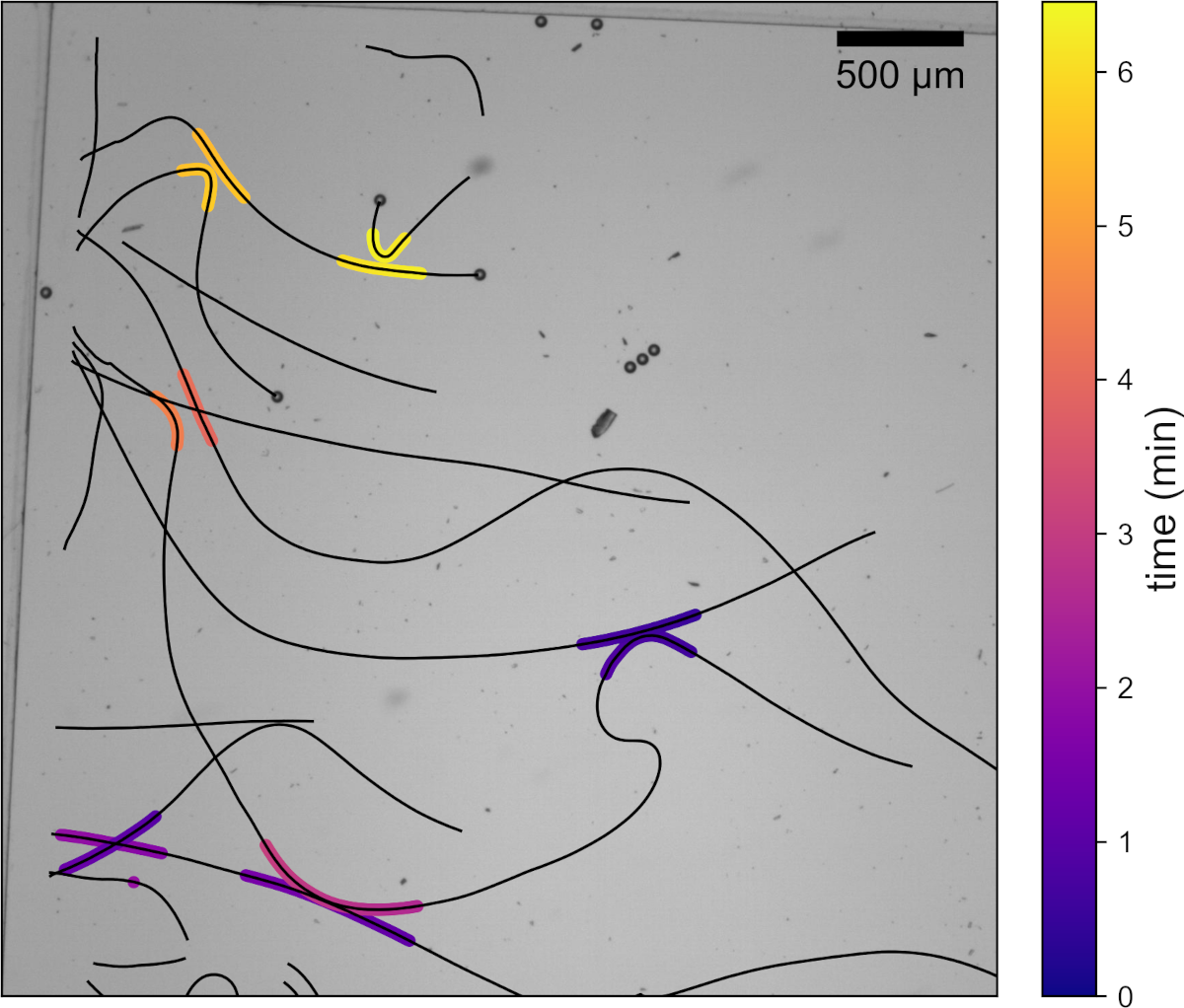}
    \caption{Identification of collision events, experiment at low number densities. Black lines: extracted trajectories, detected collision events underlaid in time-encoded color. Entrained droplets are excluded from the analysis. Background: final frame in experiment.}
    \label{SIfig:collisions}
\end{figure}

To generate experimental data on droplet-trail collisions for comparison to the CAPP model, we placed droplets at low number densities ( 0.25 to 0.50 droplets per $\rm mm^2$) in aqueous surfactant solutions in quasi-2D microfluidic reservoirs. 
We recorded and analyzed the droplet trajectories via video microscopy and mined the data from 20 experimental runs for suitable trail interactions, as follows.
Since our model assumes the preceding droplet to move in a pristine medium with isotropic chemical field, we selected interactions where its motion can locally be well approximated by a straight line (red trajectories in \figref[a,b]{fig:Autochemotactic}). Thus, we can we assume the chemical gradient in the trail to evolve according to \eqnref{eq:chemicalProfile}.
After region of interest selection, background correction and binarization, we extracted droplet coordinates via a contour search algorithm \cite{bradski2000_opencv} combined with a blob size filter for each video frame. Our model does not account for direct hydrodynamic interactions between droplets, for example entrainment. We therefore filtered the resulting coordinate set to exclude droplets whose distance within the same frame are below a set threshold of  \SI{90}{\um}, as well as droplets too close to the cell boundaries. 
Droplet trajectories were extracted from the filtered set via a Crocker-Grier type algorithm~\cite{crocker1996_methods}, providing droplet coordinates $x,y$ and speed vectors $\vec{v}$ for each recorded timestep.
We identified and analyzed interactions as follows: for each pair of trajectories, we identified matching sections where the trajectory distance fell below a threshold of \SI{220}{\um}. 
Interactions that include a trajectory endpoint were excluded as we cannot guarantee these to be complete. Segment pairs $s_1, s_2$ were sorted by time. We assume the droplet creating $s_1$ to move freely, such that $s_1$ can be safely approximated by a straight line. We note that our swimmers' dynamics are persistent Brownian rather than strictly ballistic, however, the trajectory persistence length, as seen in \figref{SIfig:collisions}, clearly exceeds the typical interaction length, so that our assumption of straight segments is reasonable. We identified the orientation $\hat{\vec{e}}_1$ of $s_1$ via a linear regression fit. If the standard deviation of this fit exceeded a certain value, the segment was considered to be too crooked, i.e. not relating to free motion, and the interaction was discarded. 
\figref{SIfig:collisions} shows trajectories from one experimental run with the numerically identified interactions marked in color.
We further discarded, by visual inspection, any interactions that were disturbed by multiple trail collisions. For segment $s_2$, we extracted the following quantities: 
\begin{enumerate}
    \item the time $\Delta t=t_2-t_1$ elapsed between the two points of closest trajectory approach, which we chose as the time delay of the interaction.
    \item for each coordinate in $s_2$, the distance to the closest point in $s_1$, i.e. distance  $d(t)$ of swimmer to trail over time. To mark crossing events, by convention, $d$ is signed via $\sgn(d) = \sgn(\vec{n}_2\times\hat{\vec{e}}_1)$, with $\vec{n}_2$ denoting the $s_2$ trajectory normal and $\times$ the 2D cross product. 
    \item from a projection of $\vec{v}_2$ on $\hat{\vec{e}}_1$, droplet speeds $v_\parallel$ parallel and $v_\perp$ perpendicular to the trail, as well as the angle $\theta$ between $\vec{v}_2$ and $\hat{\vec{e}}_1$. To avoid discontinuities in $\theta$ due to $2\pi$ periodicities, $\theta$ was calculated with respect to $-\hat{\vec{e}}_1$ if $\langle v_\parallel\rangle>0$. 
\end{enumerate}

We note that for non-specular reflections with $\alpha_0\ne0$ (\eqnref{eq:posDynamics}) the time of maximum rotation rate,  $t_\text{turn}=|\rm d\theta/d\rm t|_{\rm max}$, should be slightly delayed with respect to the time of closest approach. 

\begin{figure}
    \centering
    \includegraphics[width=\columnwidth]{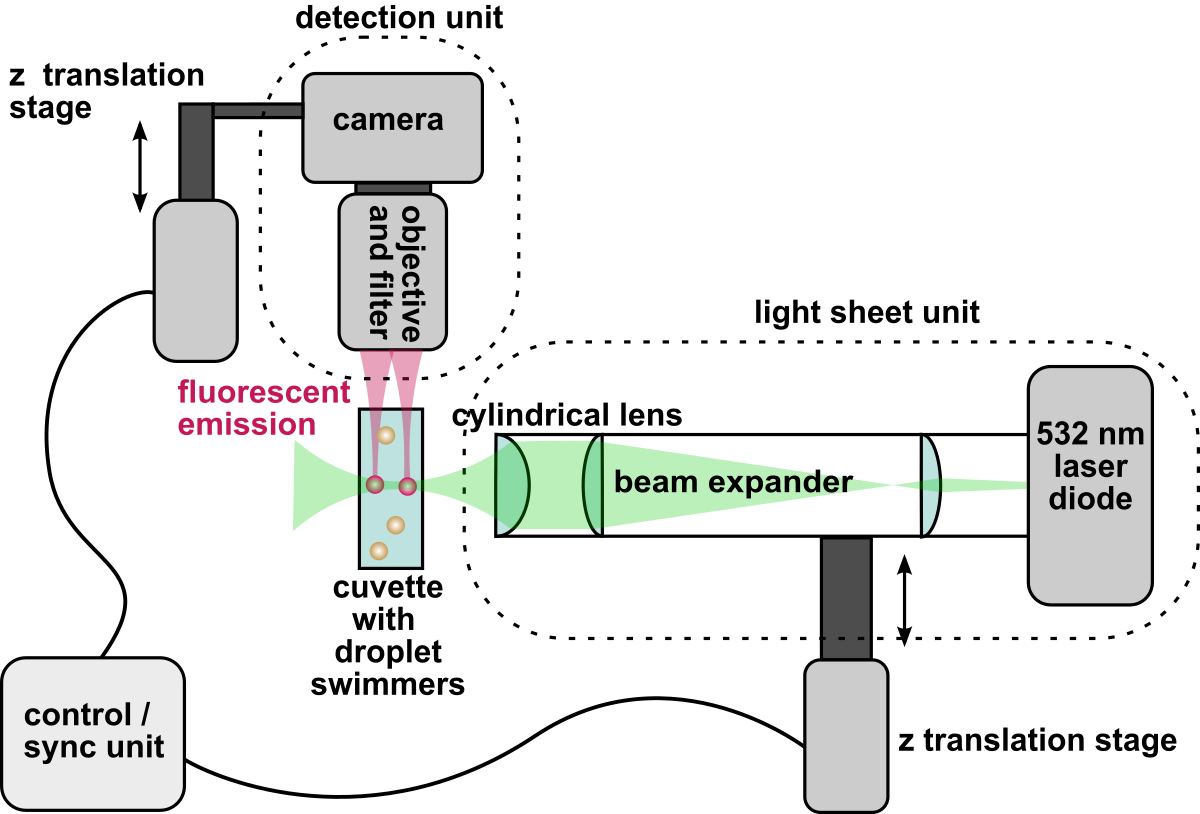}
    \caption{Fluorescent light sheet microscope, schematic.}
    \label{SIfig:lightsheet}
\end{figure}

\begin{figure}
    \centering
    \includegraphics[width=\columnwidth]{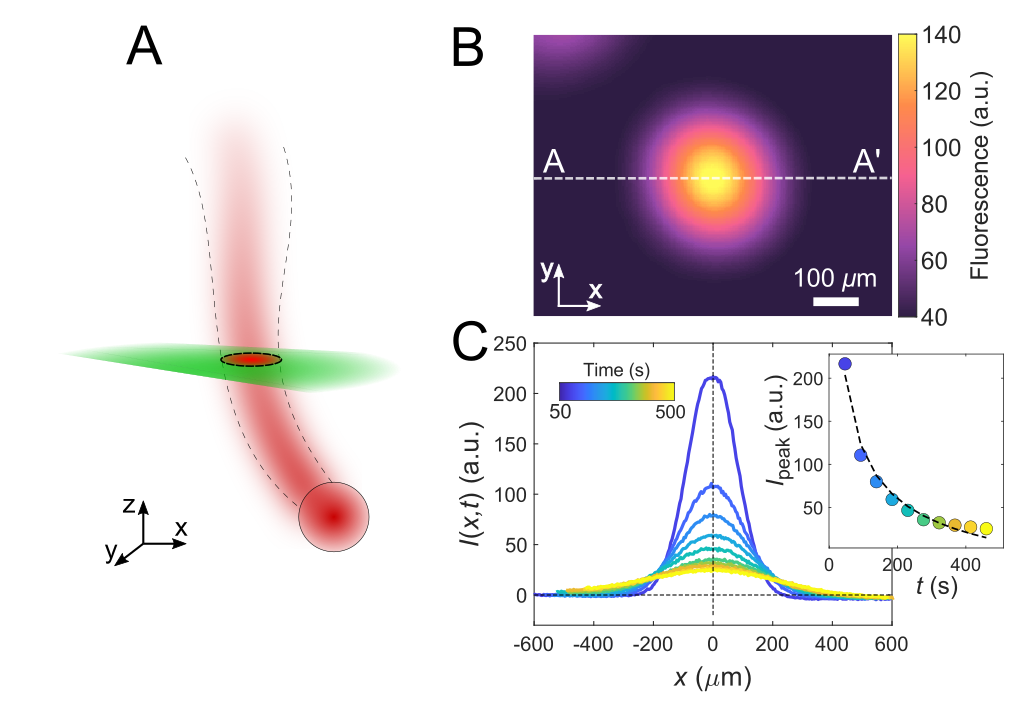}
    \caption{Visualisation of the trail in 3D. \plb{a} The schematic of trail visualisation in 3D. The green sheet represents the laser light which scans the trail at different heights obtaining images like \plb{b}. \plb{c} The temporal evolution of intensity profiles obtained by laser sheet measurements. The inset shows the peak values versus time, the dashed line is a fit to a $t^{-1/2}$ decay.}
    \label{SIfig:3DTrail}
\end{figure}

\subsection{Laser sheet fluorescent microscopy}
The light sheet /selective plane fluorescent microscopy setup (\figref{SIfig:lightsheet}) consists of an illumination unit producing the thin laser sheet and a detection unit, \textit{i.e.} camera and objective. Both units are translated vertically by synchronized $z$-stages, capturing images at a frame size of 1MP in the $xy$ plane at 4.15 $\mu$m/px resolution. Images were recorded at 150 frames per second, while the $z$-stages were driven by a sawtooth signal with an amplitude of \SI{3}{mm} and a frequency of \SI{0.7}{\hertz}, yielding a voxel size of 4.2x4.2x\SI{27}{\cubic\micro\metre}. The laser sheet's beam waist is $\approx$\SI{40}{\micro\metre} thick. Using a glass cuvette of internal size 3x\SI{3}{\square\milli\metre}, the accessible scanned sample volume is  3x3x\SI{3}{\cubic\milli\metre}.
Our samples consisted of Nile Red doped droplets with a diameter of \SI{50}{\micro\metre} in density matched mixtures of TTAB/H$_2$O/D$_2$O.

3D droplet positions were reconstructed from binarized stacks of $z$-slices for each half period of the sawtooth signal. We extracted fluorescent droplet contours for each slice \cite{bradski2000_opencv} and grouped contours associated with the same droplet within consecutive slices using a mean shift clustering algorithm \cite{pedregosa2011_scikit-learn,comaniciu2002_mean}. Time and $z$ position for each slice were calculated using timestamps provided by the camera and translation stage software interfaces, resulting in full $xyzt$ datasets. 

We also validated the diffusive spreading of the trail in 3D: we first captured the time-dependent fluorescence in the trail of a droplet sedimenting under gravity along the light sheet normal, taken at a fixed sheet position in $z$ (\figref[a]{SIfig:3DTrail}).
We extracted the fluorescence intensity profile $I(x,t)$ of a cross section of the trail along the line $\rm AA'$ (\figref[c]{SIfig:3DTrail}), as shown in the false-color sample image in \figref[b]{SIfig:3DTrail}. 
The peak values $I_{\rm peak}$ scale with $t^{-1/2}$ (inset), similar to the 2D behavior analyzed in \figref{fig:TrailVis}.

\subsection{Modeling of individual droplet-trail interactions}\label{SIsec:modelsingle}

We model the micelle trail left by a preceding droplet as a chemical field $c$ with Gaussian profile perpendicular to the direction of motion, with a width that depends on the time $\Delta t$ elapsed since the preceding droplet moved away from the point of interest. Without loss of generality we assume the preceding droplet to propel along $x$. Because the diffusion of the micelles is slow compared to the propulsion velocity of the droplets ($D_\text{fm} / (\Delta t V_0^2) \ll 1$), we can neglect the $x$-dependence of the concentration profile, as well as its time-dependence, so that the chemical field is given by the static, one-dimensional profile

\begin{equation}
c(y) = \frac{c_0}{\sqrt{4 \pi D_\text{fm} \Delta t} } \exp \left( -\frac{y^2}{4 D_\text{fm} \Delta t}\right),
\label{eq:chemicalProfile}
\end{equation}

where $D_\text{fm}$ is the diffusion coefficient of the (filled) micelles. Here, $c_0$ is a constant with units of inverse area, which can be related to the system parameters as $c_0 = \lambda / (V_0 h)$, where $\lambda$ is the rate of micelle release by the droplets, $V_0$ the droplet propulsion velocity, and $h$ the height of the Hele-Shaw cell.

We combine (\ref{eq:chemicalProfile}) with the CAPP model in Eqs.~(\ref{eq:posDynamics}--\ref{eq:axisDynamics}) of the main text, and further ignore translational and rotational diffusion as they are both negligible within the short time scale of a single interaction event. We thus solve the following deterministic equations for the incoming droplet's position $(x,y)$ and orientation $\theta$, which is defined as the angle between $\bm{n}$ and the $x$-axis (see \figref[d]{fig:Autochemotactic}):

\begin{align}
\dot{x} &= V_0 \cos \theta \label{eq:xDyn} \\
\dot{y} &= V_0 \sin \theta +\frac{\alpha c_0 y}{\sqrt{2\pi} (2 D_\text{fm} \Delta t)^{3/2}} \exp \left( -\frac{y^2}{4 D_\text{fm} \Delta t}\right)   \label{eq:yDyn} \\
\dot{\theta} &= -\frac{\Omega c_0 y}{\sqrt{2\pi} (2 D_\text{fm} \Delta t)^{3/2}} \exp \left( -\frac{y^2}{4 D_\text{fm} \Delta t} \right) \cos \theta. \label{eq:thetaDyn}
\end{align}

\subsection{Fitting of individual droplet-trail interactions}\label{SIsec:fitting}

For the diffusion coefficient of the micelles, we used the measured value $D_\text{fm} = \SI{52.5}{\square\um\per\second}$. The droplet velocity $V_0$ was estimated by the average velocity in the given trajectory, and the experimentally-measured time lag $\Delta t$ was used as an input. Initial conditions for the time evolution of $(x(t),y(t),\theta(t))$ were obtained from the data as described in the previous section: $y(0) \approx \SI{200}{\um}$ corresponds to the initial value of the signed distance $d_\mathrm{max}$, $\theta(0)$ to the incidence angle, and $x(0)=0$ without loss of generality. Thus, the only unknown parameters are the two products $\Omega c_0$ and $\alpha c_0$, which are used as fit parameters.

\begin{figure*}
    \centering
    \includegraphics[width=\textwidth]{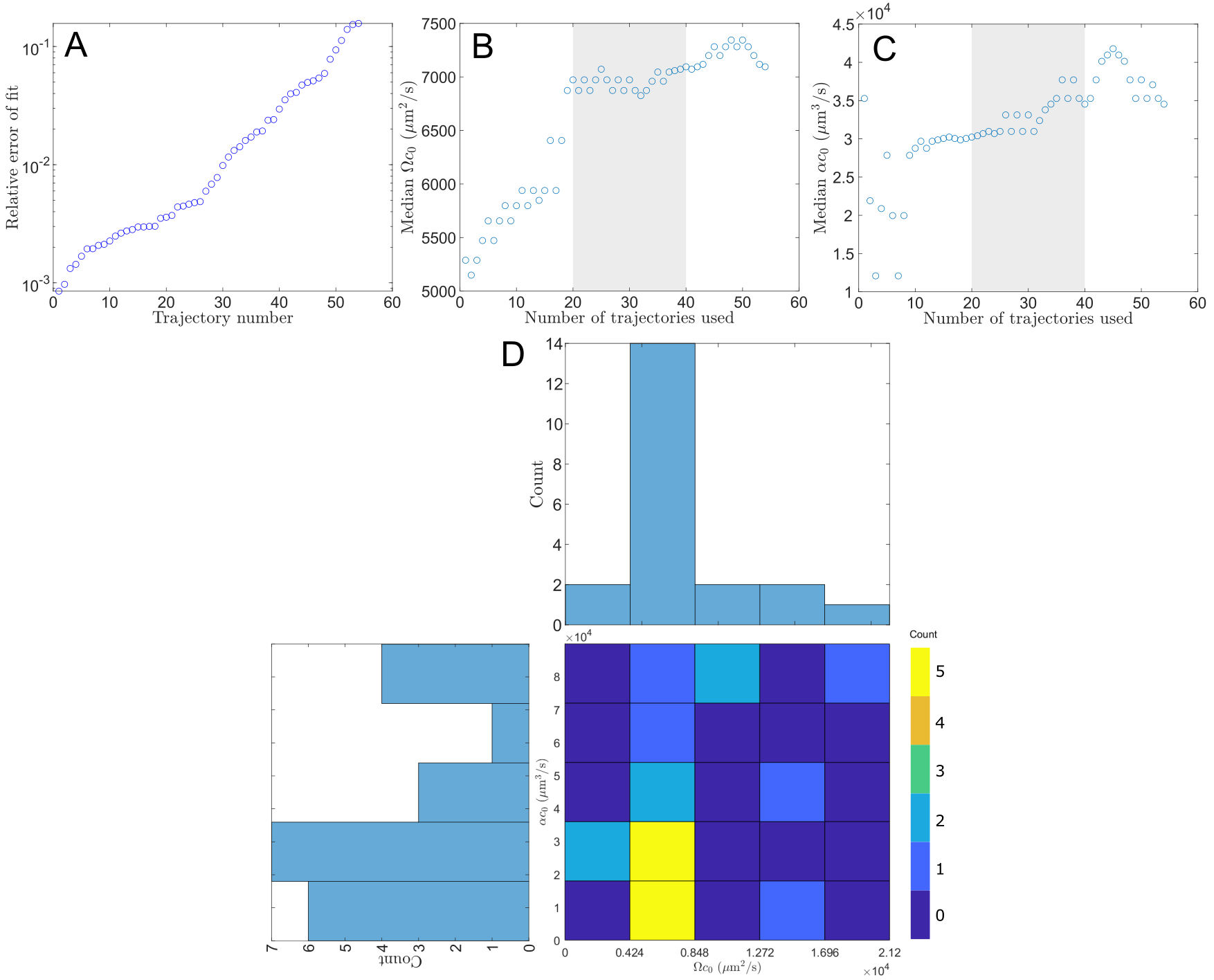}
    \caption{Analysis of the fits to the 54 experimental reflection events with large incidence angle $\theta_\mathrm{inc}>60^\circ$, using two fitting parameters $\Omega c_0$ and $\alpha c_0$. (a) The fits are ordered from best to worst according to the error of the fit. (b,c) Median values of (b) $\Omega c_0$ and (c) $\alpha c_0$ as a function of the the number $n$ of trajectories used in the calculation of the median value. We observe a plateau at $\Omega c_0 \approx  7 \cdot 10^3$ $\mu$m$^2/$s and $\alpha c_0 \approx  3 \cdot 10^4$ $\mu$m$^3/$s (gray band), when a sufficient number of trajectories is considered ($n \gtrsim 20$) but the worst fits are left out ($n \lesssim 40$). (d) As an example, we show the histogram for the values of $\Omega c_0$ and $\alpha c_0$ (two-dimensional, as well as projected along each dimension) when using the $n=21$ best fits, which corresponds to using only the fits with relative error $<4 \cdot 10^{-3}$. }
    \label{SIfig:analysisfits}
\end{figure*}

For our analysis, we focused on sharp reflection events with $\theta_\mathrm{inc}>60^\circ$, for which our minimal model of the trail as a static Gaussian with $x$-independent width is best justified (for low incidence angles, interactions occur over longer times and a wider range of $x$-values, which may affect the validity of our approximation and introduce differences between parallel and antiparallel reflection events). The 54 reflection events were fitted and ordered from best to worst fit according to the fit error. By analyzing the median values of $\Omega c_0$ and $\alpha c_0$ calculated from the $n$ best fits as a function of $n$, we found that the median values stabilize at $\Omega c_0 \approx \SI{7e3}{\square\um\per\second}$ and $\alpha c_0 \approx \SI{3e4}{\um^3\per\second}$ between $n \approx 20$ and $n \approx 40$. Smaller $n$ values are susceptible to the noise due to small number statistics, whereas for larger $n$ we would include bad fits that skew the distribution, presumably corresponding to non-ideal trajectories (e.g.~interactions with curved trails or measurement artefacts due to global drift in the chamber). Details can be found in Fig.~\ref{SIfig:analysisfits}.
We note that we do not distinguish between parallel and antiparallel interactions: for $\theta_\text{inc}>90^{\circ}$, we convert to $180^\circ - \theta_\text{inc}$.

\subsection{Modeling of collective behavior}\label{SIsec:modelmany}

To model the behavior of a many-droplet system with $N$ droplets, we use the equations of the CAPP model, Eqs.~(\ref{eq:posDynamics}--\ref{eq:axisDynamics}) of the main text with $i=1,...,N$, together with an evolution equation for the concentration field $c(\bm{r},t)$ of the filled micelles
\begin{equation}
\partial_t c(\bm{r},t) = D_\mathrm{fm} \nabla^2 c + \lambda \sum_{j=1}^N \delta(\bm{r}-\bm{r}_j)
\label{eq:solDynamics}
\end{equation}
where $\lambda$ is the rate of micelle production by the droplets, which are modeled as point sources. Eq.~(\ref{eq:solDynamics}) can be solved by considering a superposition of instant point sources released at all times $t'<t$, at locations $\bm{r}_j(t')$ for all $j=1,...,N$, each of which undergoes two-dimensional diffusion:
\begin{equation}
c(\bm{r},t) =  \frac{\lambda}{h} \sum_{j=1}^N \int_{-\infty}^t \mathrm{d}t' \frac{\exp \left( - \frac{(\bm{r}-\bm{r}_j(t'))^2}{4D_\mathrm{fm}(t-t')} \right)}{4\pi D_\mathrm{fm}(t-t')} 
\label{eq:solute}
\end{equation}
Its gradient can be directly calculated as
\begin{equation}
\nabla c(\bm{r},t) = -2  \frac{\lambda}{h} \sum_{j=1}^N \int_{-\infty}^t \mathrm{d}t' \frac{[\bm{r}-\bm{r}_j(t')]\exp \left( - \frac{(\bm{r}-\bm{r}_j(t'))^2}{4D_\mathrm{fm}(t-t')} \right)}{\pi[4 D_\mathrm{fm}(t-t')]^2} 
\label{eq:gradsolute}
\end{equation}

The combination of (\ref{eq:posDynamics}--\ref{eq:axisDynamics}) in the main text with (\ref{eq:gradsolute}) gives a closed set of evolution equations for the system. However, as a consequence of assuming the particles to be point sources and to respond to the local gradient, this model leads to spurious, infinitely-strong self-interactions of each droplet with the micelles it just released. These can in principle be regularized by considering the finite size of droplets. Nevertheless, because we are interested in describing collective behavior in systems with a large number of particles, we can side-step this problem by simply neglecting the self-interaction and defining the local gradient that enters Eqs.~(\ref{eq:posDynamics}--\ref{eq:axisDynamics}) as
\begin{equation}
\nabla c|_{\bm{r}=\bm{r}_i} = -2  \frac{\lambda}{h} \sum_{\substack{j=1 \\ j\neq i}}^N \int_{-\infty}^t \mathrm{d}t' \frac{[\bm{r}_i-\bm{r}_j(t')]\exp \left( - \frac{(\bm{r}_i-\bm{r}_j(t'))^2}{4D_\mathrm{fm}(t-t')} \right)}{\pi[4 D_\mathrm{fm}(t-t')]^2} 
\label{eq:gradsolute2}
\end{equation}
Using (\ref{eq:gradsolute2}), we numerically solve equations (\ref{eq:posDynamics}--\ref{eq:axisDynamics}) using Brownian dynamics using the parameters $V_0$, $D_\mathrm{fm}$, $D_t$ and $D_r$ as measured experimentally, and the parameters $\Omega c_0$ and $\alpha c_0$ as extracted from the fits to individual interaction events. The simulations use an Euler time step of $10^{-2}$~s, and a square box with periodic boundary conditions with minimum-image convention for all distances $\bm{r}_i-\bm{r}_j(t')$ in (\ref{eq:gradsolute2}). The results in Fig.~\ref{fig:2DCaging}d of the main text used a box of side length $L=6.32$~mm and particle numbers $N=1,72,148,296,343$, resulting in the same particle area densities as in the experiments, and the trajectory in Fig.~\ref{fig:2DCaging}d corresponds to the simulation with $L=6.32$~mm and $N=343$.

In the experimental data, since ballistic runs are uncorrelated, the long-time dynamics presumably correspond to a diffusive random walk, however, it is not feasible to reliably quantify the respective exponent in the experimental MSD due to the limited lifetime of the droplets and the finite size of the experimental cell.
\onecolumngrid


\section{Supplemental Videos}

\begin{center}
    \includegraphics[width=.58\columnwidth]{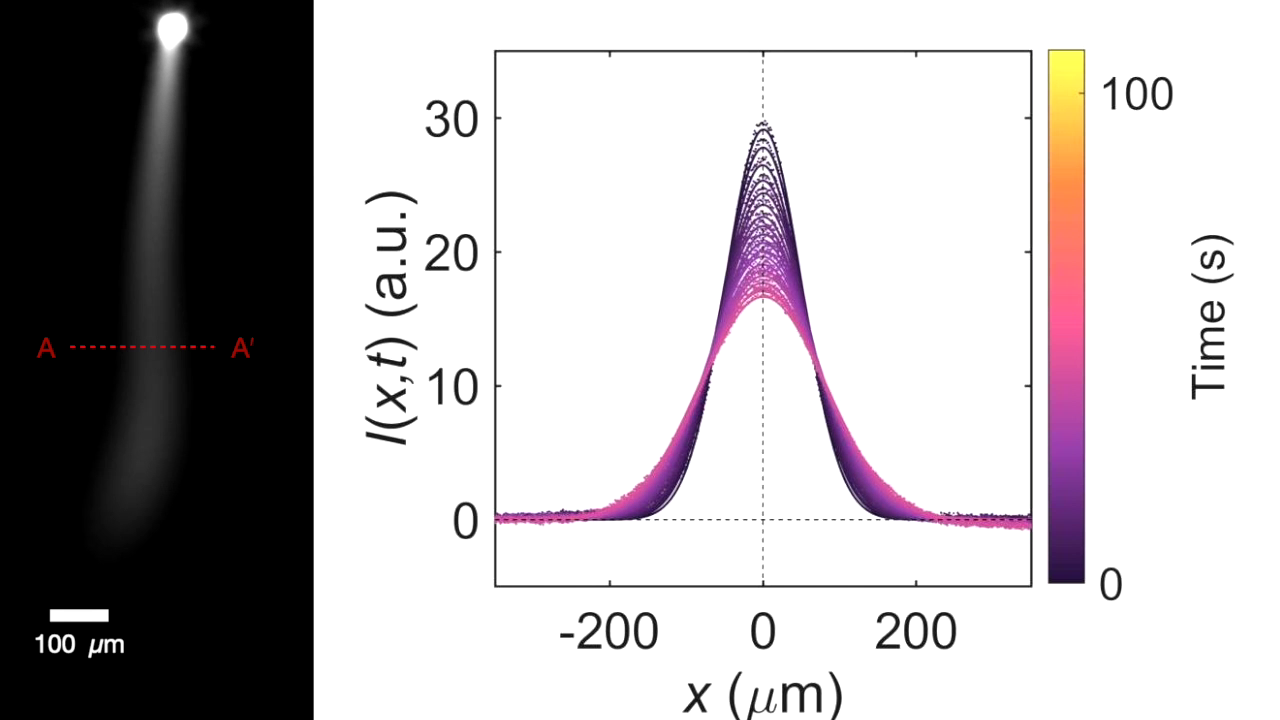}
\end{center}

\textbf{Supplemental Video S1:} Quantitative imaging of the chemical trail of a droplet via fluorescence microscopy. The animated plot visualizes the fluorescence measured at the same time along the line AA' in the video on the left. Video sped up 10x. The swimming medium is 5 wt.$\%$ TTAB solution.

\begin{center}
    \includegraphics[width=.58\columnwidth]{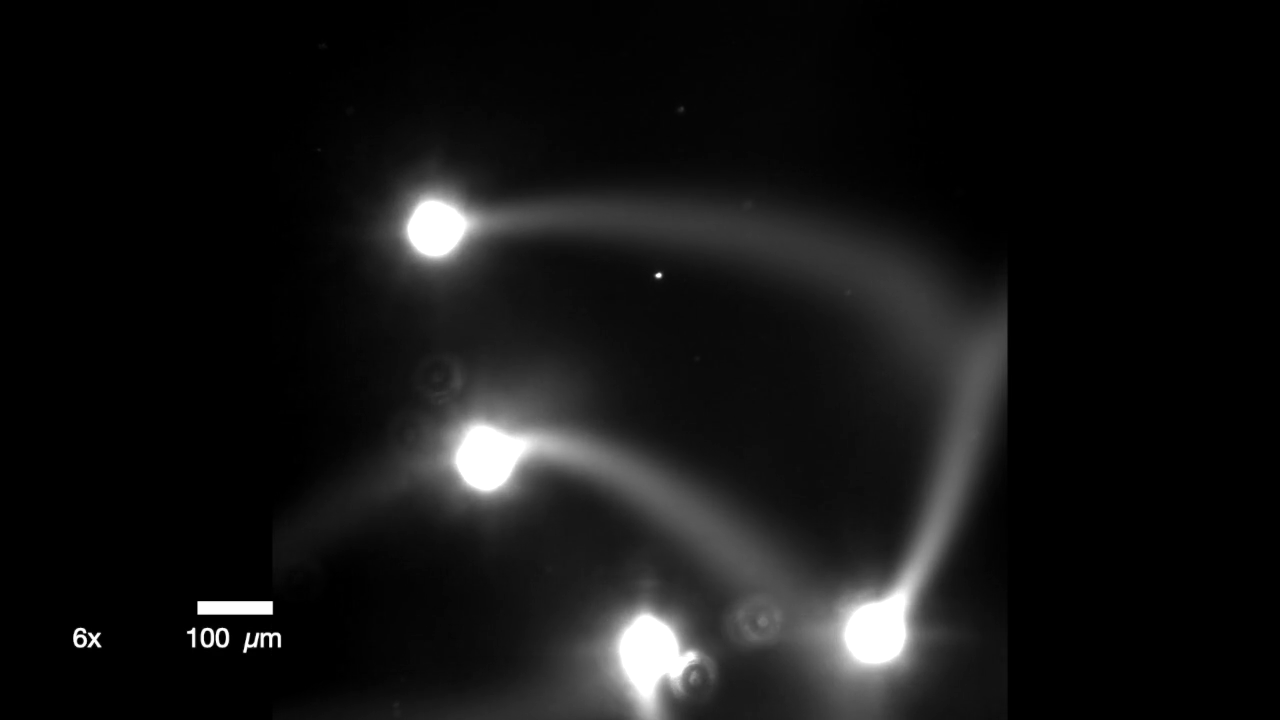}
\end{center}

\textbf{Supplemental Video S2:} Autochemotactic interaction between a droplet and a trail: Crossing events. Fluorescent videomicroscopy of NileRed dyed droplet swimmers, with the dye comigrating into the swimmer trails. The swimming medium is 5 wt.$\%$ TTAB solution.

\begin{center}
    \includegraphics[width=.58\columnwidth]{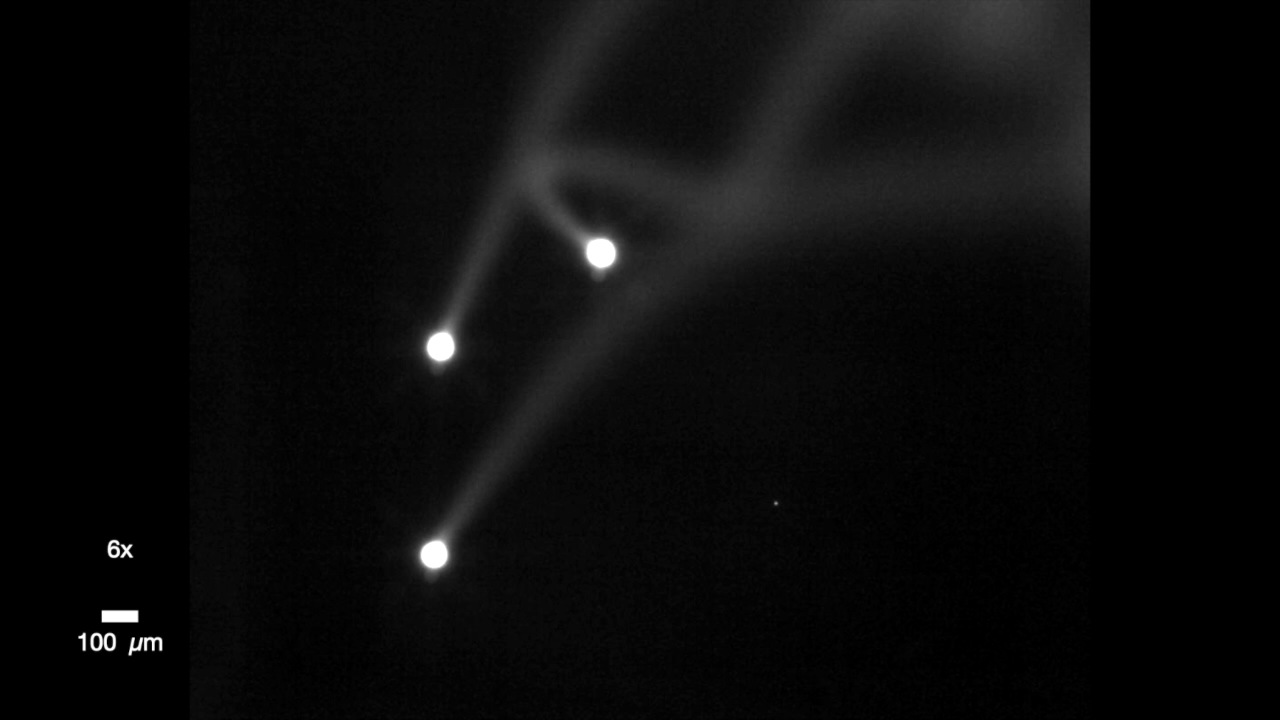}
\end{center}

\textbf{Supplemental Video S3:} Autochemotactic interaction between a droplet and a trail: Reflecting events. The illumination is changed back and forth between the fluorescence and visible light. The swimming medium is 5 wt.$\%$ TTAB solution.

\begin{center}
    \includegraphics[width=.58\columnwidth]{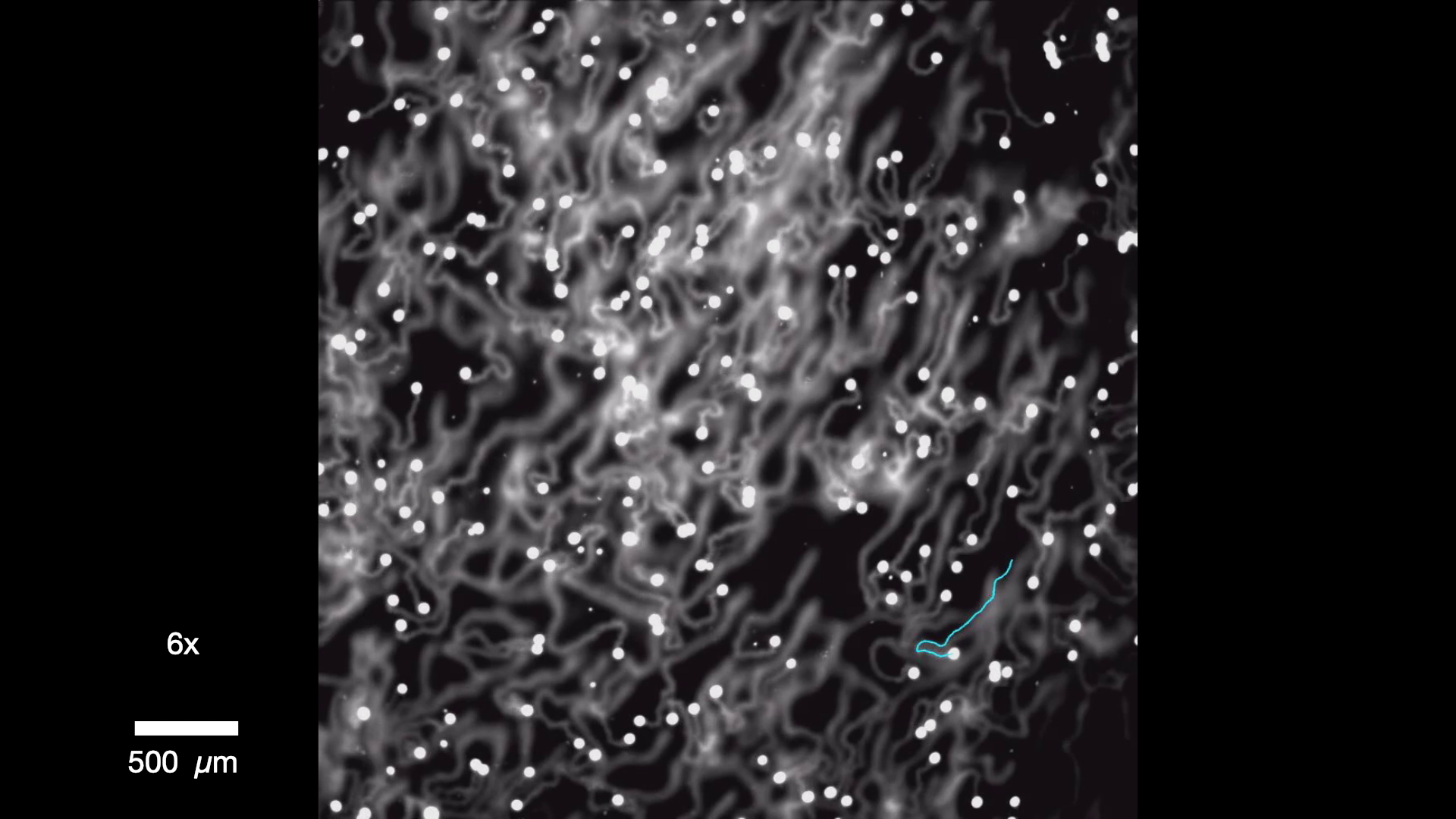}
\end{center}

\textbf{Supplemental Video S4:} 2D autochemotactic caging in a Hele-Shaw (quasi-2D) cell, observed under fluorescent microscopy. The TTAB concentration was increased to 25 wt.$\%$ compared to the experiments used for quantitative analysis (5 wt.$\%$), for better trail visibility. The less persistent swimming is due to the higher solubilization rate. Stills from this video were used in Fig. 3.

\begin{center}
    \includegraphics[width=.58\columnwidth]{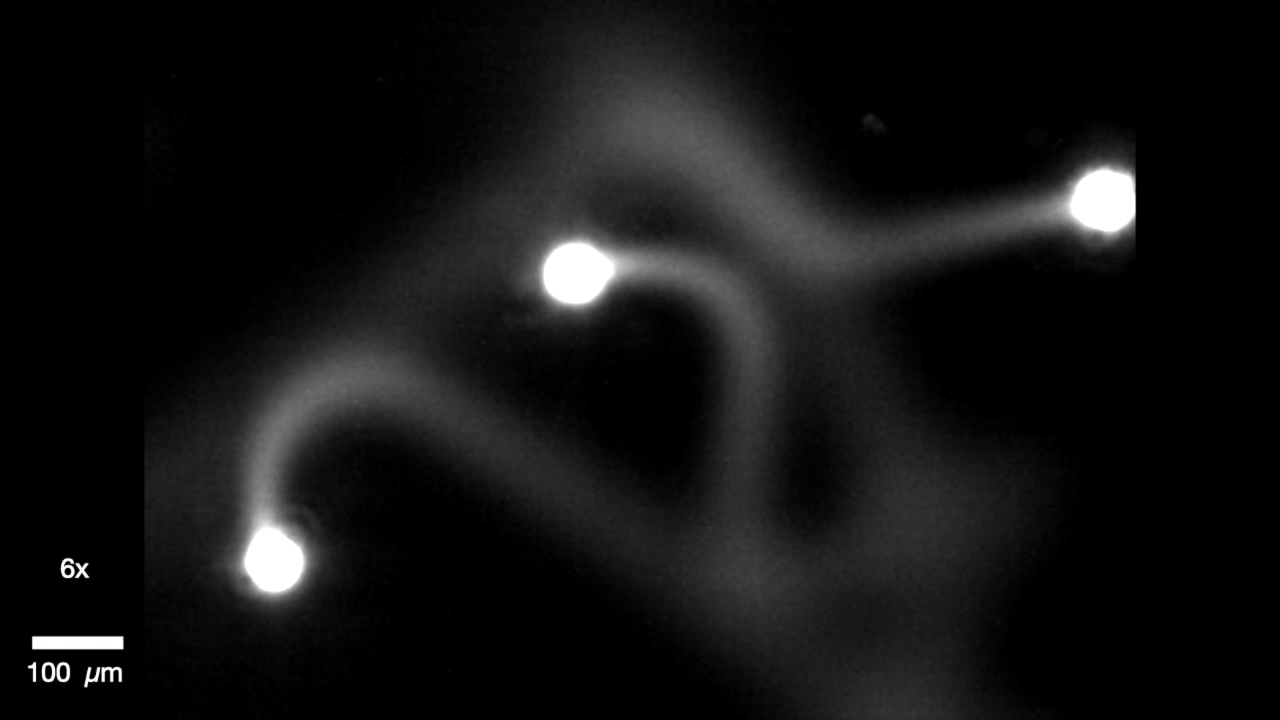}
\end{center}

\textbf{Supplemental Video S5:} An example caging event in a \textit{dilute} active emulsion followed by a “cage escape”, observed under fluorescence microscopy. The swimming medium is 5 wt.$\%$ TTAB solution.

\begin{center}
    \includegraphics[width=.58\columnwidth]{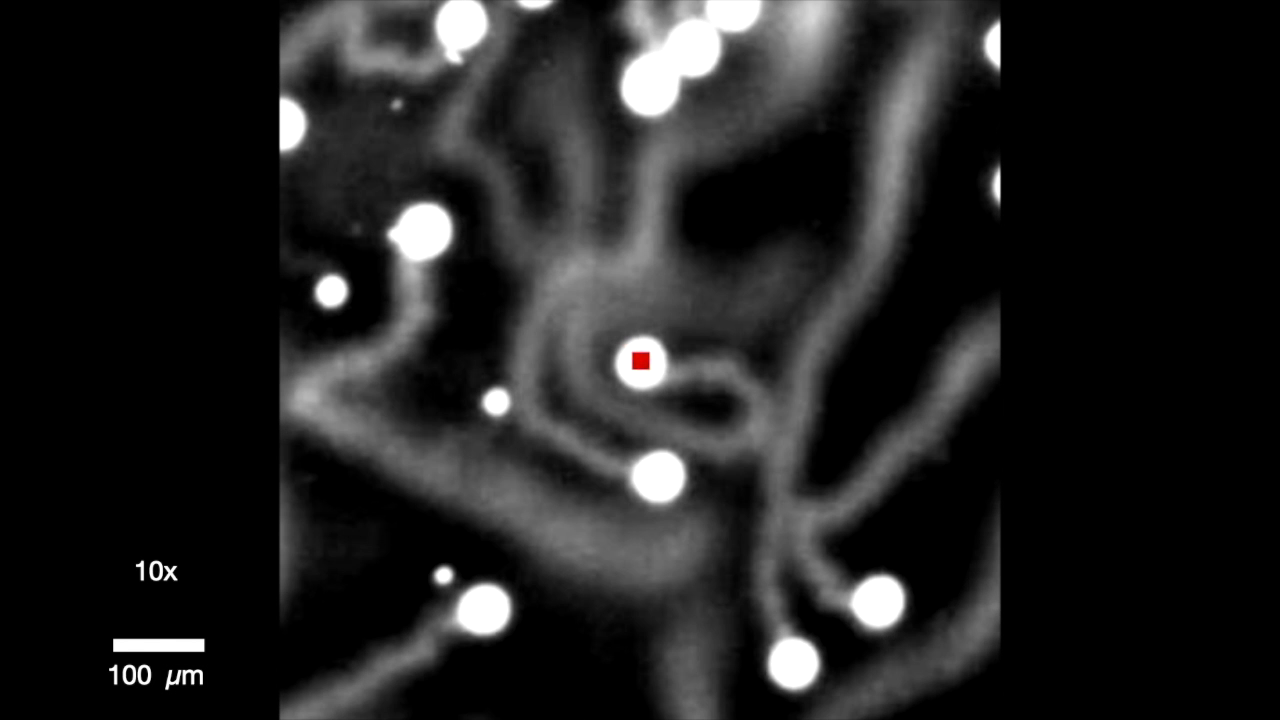}
\end{center}

\textbf{Supplemental Video S6:} Cage escape due to the chemical buildup within a \textit{dense} active emulsion, observed under fluorescence microscopy. The droplet, marked red, is tracked and centered in the movie frames. The swimming medium is 25 wt.$\%$ TTAB solution.

%

\end{document}